\documentclass[usenatbib,useAMS]{mn2e}
\usepackage{amssymb,amsbsy}
\usepackage{graphicx}
\usepackage{bm}
\usepackage{color}
\def\apj{ApJ}
\def\apjs{ApJS}

\def\apjl{ApJL}
\def\nat{Nat}
\def\araa{ARA\&A}
\def\pasa{Proc. Astron. Soc. Australia}

\def\aap{A\&A}
\def\zap{Zeitschrift f\"ur Astrophysik}
\def\mnras{MNRAS}
\def\jfm{J. Fluid Mech.}
\def\prl{Phys. Rev. Lett.}
\def\pre{Phys. Rev. E}
\def\jcp{J. Comp. Phys.}
\def\lsim{~\raise0.3ex\hbox{$<$}\kern-0.75em{\lower0.65ex\hbox{$\sim$}}~}
\def\gsim{~\raise0.3ex\hbox{$>$}\kern-0.75em{\lower0.65ex\hbox{$\sim$}}~}

\title[The turbulent origin of Larson's laws]{A supersonic turbulence origin of Larson's laws}

\author[A. G. Kritsuk, C. T. Lee and M. L. Norman]%
{Alexei G. Kritsuk,$^{1}$\thanks{E-mail: akritsuk@ucsd.edu} 
Christoph T. Lee$^{1}$
and Michael L. Norman$^{1,2}$\\
$^{1}$Physics Department and CASS, University of California, San Diego, 9500 Gilman Drive, La Jolla, CA 92093-0424, USA\\
$^{2}$San Diego Supercomputer Center, University of California, San Diego,  10100 Hopkins Drive, La Jolla, CA 92093-0505, USA}

\begin{document}

\date{Accepted 2013 September 23. Received 2013 September 23; in original form 2013 January 9}

\pagerange{\pageref{firstpage}--\pageref{lastpage}} \pubyear{2013}

\maketitle

\label{firstpage}

\begin{abstract} 
We revisit the origin of Larson's scaling laws describing the structure and kinematics of molecular clouds.
Our analysis is based on recent observational measurements and data from a suite of six  
simulations of the interstellar medium, including effects of self-gravity, turbulence, magnetic field and multiphase 
thermodynamics. Simulations of isothermal supersonic turbulence reproduce observed slopes in 
linewidth--size and mass--size relations. Whether or not self-gravity is included, the linewidth--size
relation remains the same. The mass--size relation, instead, substantially flattens below the sonic 
scale, as prestellar cores start to form. Our multiphase models with magnetic field 
and domain size $200$~pc reproduce both scaling and normalization of the first Larson law. 
The simulations support a turbulent interpretation of Larson's relations. This interpretation 
implies that: (i) the slopes of linewidth--size and mass--size correlations are determined by the inertial 
cascade; (ii) none of the three Larson laws is fundamental; (iii) instead, if one is known, the other two 
follow from scale invariance of the kinetic energy transfer rate. It {\em does not} imply that gravity
is dynamically unimportant. The self-similarity of structure established by the turbulence 
breaks in star-forming clouds due to the development of gravitational instability in the vicinity of 
the sonic scale. The instability leads to the formation of prestellar cores with the characteristic 
mass set by the sonic scale. The high-end slope of the core mass function predicted by the 
scaling relations is consistent with the Salpeter power-law index. 
\end{abstract}

\begin{keywords}
turbulence --
methods: numerical --
stars: formation --
ISM: structure.
\end{keywords}

\section{Introduction}\label{intro}
Understanding physics underlying structure formation and evolution of molecular clouds (MCs) is an important stepping 
stone to a predictive statistical theory of star formation \citep{mckee.07}. 
Statistics of non-linear density, velocity, and magnetic field fluctuations in MCs may have imprints in the star formation rate 
and the stellar initial mass function \citep{krumholz.05,hennebelle.11,padoan.11,padoan.02,padoan....07,hennebelle.08}. 
Non-linear coupling of self-gravity, turbulence and magnetic 
field in MCs is believed to regulate star formation \citep{maclow.04}. But how does this actually work? Answering this question
would ultimately help one to break new ground for ab initio star formation simulations. This paper mostly deals with one 
particular aspect of the cloud structure formation, namely the interaction of turbulence and gravity, by confronting observations 
with numerical simulations and theory.

\citet{larson81} established that for many MCs their internal velocity dispersion, $\sigma_{u}$,
is well correlated with the cloud size, $L$, and mass, $m$. Since the power-law form of the correlation,
$\sigma_{u}\propto{}L^{0.38}$, and the power index, $0.38\sim1/3$, were similar to those of the 
\citet[][hereafter K41]{kolmogorov41a,kolmogorov41b} turbulence, he suggested that observed non-thermal 
linewidths may originate from a `common hierarchy of interstellar turbulent motions'. The clouds would also 
{\em appear} mostly gravitationally bound and in approximate virial equilibrium, as there was a close positive 
correlation between their velocity dispersion and mass, $\sigma_{u}\propto{}m^{0.20}$. However, Larson suggested 
that these structures `cannot have formed by simple gravitational collapse' and should be at least partly created 
by supersonic turbulence. This seminal paper preconceived many important ideas in the field and
strongly influenced its development for the past 30 years.

\citet{myers83} studied 43 smaller dark clouds and confirmed the existence of significant linewidth--size and density--size correlations
found earlier by Larson for larger MCs. He acknowledged that two distinct interpretations of the data are possible: (i) the
linewidth-size relation also known as Larson's first law ($\sigma_u\propto R^{0.5}$) arises from a Kolmogorov-like cascade 
of turbulent energy; (ii) the same relation results from the tendency of clouds to reside in virial equilibrium ($\sigma_u\propto \rho^{0.5}R$,
Larson's second law) and from the tendency of the mean cloud density to scale inversely linearly with the cloud size ($\rho\propto R^{-1}$,
Larson's third law). He concluded that the available data `do not permit a clear choice between these interpretations'.

\citet[][hereafter SRBY87]{solomon...87} confirmed Larson's study using observations of $^{12}$CO emission with improved 
sensitivity for a more homogeneous sample of 273 nearby clouds. Their linewidth--size 
relation, $\sigma_{u}=1.0\pm0.1S^{0.5\pm0.05}$~km~s$^{-1}$, however, had a substantially steeper slope
than Larson's, more reminiscent of that for clouds in virial equilibrium,\footnote{We deliberately keep the original 
notation used by different authors for the cloud size (e.g. 
the size parameter in parsecs, $S=D\tan(\sqrt{\sigma_l\sigma_b})$; 
the maximum projected linear extent, $L$;
the radius, $R=\sqrt{A/\pi}$, defined for a circle with area, $A$, equivalent to that of cloud) 
to emphasize ambiguity and large systematic errors in the cloud size and mass estimates due 
to possible line-of-sight confusion, ad hoc cloud boundary definitions \citep{heyer...09}, and various X-factors
involved in conversion of a tracer surface brightness into the H$_2$ column density.}   
\begin{equation}
\sigma_{u}=\left(\pi{}G\Sigma\right)^{1/2}R^{1/2},
\label{vir}
\end{equation}
partly because the SRBY87 clouds had approximately constant molecular gas surface density 
($\Sigma\propto\rho R\approx {\rm const.}$) independent of their size. The surface density--size 
relation -- another form of Larson's third law -- can be derived by eliminating $\sigma_u$ from his first two 
relations: $\Sigma\propto{}mL^{-2}\propto{}L^{0.38/0.20-2}\propto{}L^{-0.1}$. Thus any two of Larson's laws
imply the other, leaving open a question of which two laws, if any, are actually fundamental.

Whether or not the third law simply reflects observational selection effects stemming from limitations in 
dynamic range of available observations is still a matter of debate \citep{larson81,kegel89,scalo90,ballesteros.02,schneider.04}.
\citet*{lombardi..10b} studied the structure of nearby MCs by mapping the dust column density that is believed to be 
a robust tracer for the molecular hydrogen \citep[cf.][]{padoan.....06}. The dust extinction maps usually have 
a larger (although not much larger) dynamic range (typically $\gsim10^2$) than that feasible with simple 
molecular tracers. \citet{lombardi..10b} verified Larson's third law  ($\Sigma\approx {\rm const.}$) with a very small 
scatter for their MCs defined with a given extinction threshold. This trivial result is determined by how the
size of a cloud or clump is defined observationally and is physically insignificant \citep*[see also][]{ballesteros..12}. 
\citet{lombardi..10b} also demonstrated that the mass--size version ($m\propto R^2$) of the third law applied 
within a single cloud using different thresholds does not hold, indicating that individual clouds cannot be 
described as objects characterized by constant column density. Three-dimensional numerical simulations that 
are free of such selection effects nevertheless reproduce a robust mass--size correlation for density structures 
in supersonic turbulence with a slope of $2.3-2.4$, i.e. steeper than $m\propto R^2$ \citep*{ostriker..01,kritsuk...07a,kritsuk...09}. 
Thus simulations do not support the constant average column density hypothesis and qualitatively agree 
with single-cloud results of \citet{lombardi..10b}.

Assuming that $\Sigma={\rm const.}$ for all clouds, SRBY87 evaluated the `X-factor' to convert the luminosity in 
$^{12}$CO~$(1-0)$ line to the MC mass. It would seem that the new power index value $\sim0.5$ ruled out 
Larson's hypothesis that the correlation reflects the Kolmogorov law. In the absence 
of  robust predictions for the velocity scaling in supersonic turbulence \citep[cf.][]{passot..88}, a
simple virial equilibrium-based interpretation of linewidth--size relation appealed to many in the 1980s.
The central supporting observational evidence for a gravitational origin of Larson's relations has been 
the scaling coefficient in the first law, which coincided with the predicted value for gravitationally bound clouds 
$\sim\sqrt{\pi G\Sigma}$. 
In particular, the observed gravitational energy of giant molecular clouds (GMCs) in the outer Galaxy with the total molecular 
mass in excess of $\sim10^4$~M$_{\odot}$ was found comparable to their observed kinetic energy \citep*{heyer..01}.

Since then views on this subject have remained polarized. For instance, \citet{ballesteros....11,ballesteros.....11} 
argue that MCs are in a state of `hierarchical and chaotic gravitational collapse', while \citet*{dobbs..11} 
believe that GMCs are `predominantly gravitationally unbound objects'. As some authors tend to deny 
the existence of a unique linewidth--size relation outright, acknowledging the possibility that different 
clouds may exhibit different trends \citep[e.g.][]{ballesteros....11}, others argue that it is difficult to 
identify the relation using samples that have too small a dynamic range. In addition, the scaling laws
based on observations of different tracers may differ as they probe different density regimes and 
different physical structures \citep{hennebelle.12}.

This persistent dualism in interpretation largely stems from the fundamentally limited nature of information that can 
be extracted from observations (e.g. due to low resolution, projection effects, wide 
variation in the CO-H$_2$ conversion factor, active feedback in regions of massive star formation, etc.) 
\citep{shetty..10} and from our limited understanding of supersonic turbulence \citep[e.g.][]{ballesteros...07}. 
Early on, supersonic motions were thought to be 
too dissipative to sustain a Kolmogorov-like cascade. Later, numerical simulations showed that direct 
dissipation in shocks constitutes a small fraction of total dissipation and that turbulent velocities have the 
Kolmogorov scaling \citep*{passot..88,porter..94,falgarone....94}. This result based on simulations at Mach 
numbers around unity left the important role of strong density fluctuations underexposed \citep[cf.][]{fleck83,fleck96} 
up until simulations at higher Mach numbers demonstrated that the velocity scaling is in fact non-universal \citep{kritsuk...06}.

The key point of contention is the question of whether the hierarchically structured clouds represent 
quasi-static bound objects in virial {\em equilibrium} over approximately  four decades in linear scale 
\citep[e.g. SRBY87;][]{chieze87} or instead they represent {\em non-equlibrium} dissipative structures 
borne in the self-gravitating turbulent interstellar medium \citep[ISM; e.g.][]{fleck83,elmegreen.04,maclow.04,lequeux05,ballesteros...07,hennebelle.12}. 
While both original interpretations of Larson's laws adopt the same form of the Navier--Stokes (N-S)
equation as the starting point of their analyses, distinct conclusions for the origin of the
observed correlations are ultimately drawn. 

Note that the two interpretations are mutually exclusive in the presence of large-scale turbulence with 
the energy injection length-scale of the order of the Galactic molecular disc thickness.
This description of the ISM turbulence is supported by the extent of the observed linewidth--size and 
mass--size correlations which continue up to $\sim100$~pc with no sign of flattening 
\citep{hennebelle.12} as well as by simulations of supernova-driven ISM 
turbulence, which suggest the integral scale of  $75-100$~pc \citep{joung.06,deavillez.07,gent....12}. 
There have been scenarios, where self-gravitating clouds are instead microturbulent \citep[e.g.][]{white77}. 
For such clouds and their substructure, virial analysis would be rigorously justified down to the 
integral scale of microturbulence \citep{chandrasekhar51,bonazzola....87,bonazzola.....92,schmidt.11}. 
If turbulence were driven within clouds by their own collapse, virial analysis would be applicable to
the whole clouds only.
In this case, however, turbulent velocities would originate from and get amplified by gravitational collapse 
of the cloud \citep[e.g.][]{henriksen.84,zinnecker84,biglari.88,henriksen91}. Hence one would naively
expect $\sigma_u\propto R^{-1/2}$, which contradicts the observed linewidth--size relation \citep{robertson.12}.

In essence, the discrepancy between turbulent and virial interpretations can be tracked back to the 
ansatz, used to derive the virial theorem from the N-S equation in the form of the Lagrange identity, 
\begin{equation}
2(E_{\rm kin}+E_{\rm int})+E_{\rm mag}-E_{\rm grav}=0,\label{li}
\end{equation}
routinely applied to MCs \citep[e.g.][]{mckee.92,ballesteros06}. 
This form is incompatible with the basic nature of the direct inertial cascade as it ignores 
the momentum density flux across the fixed Eulerian cloud boundary that is generally of the 
same order as the volume terms \citep[e.g.][]{dib....07}. 
Numerical simulations show that in star-forming clouds large turbulent pressure does 
not provide mean support against gravity. 
Instead, turbulent support in self-gravitating media is on average negative, i.e. promoting the collapse 
rather than resisting it \citep{klessen00,schmidt..12}. 
Thus neither virial balance condition~(\ref{vir}) nor energy equipartition condition $E_{\rm kin}(R)\sim E_{\rm grav}(R)$
can actually guarantee boundedness of clouds embedded in the general field of large-scale turbulence.
Since virial conditions~(\ref{vir}) and (\ref{li}) do not comply with the N-S equation in the 
presence of large-scale turbulence, the fact that some of the most massive GMCs follow the first Larson 
law with a coefficient $\sim\sqrt{\pi G\Sigma}$ can be merely coincidental. This does not rule out a role for 
gravity in MC dynamics, as otherwise the clouds would never form stars or stellar clusters.

Recent progress in  observations, theory and numerical simulations of ISM turbulence prompts us to revisit the 
origin of Larson's relations. In Section~\ref{sim}, we describe the simulations that
will be used in Section~\ref{lws} to give a computational perspective on the origin of the first Larson law.
Section~\ref{ms} continues the discussion with the mass--size relation.
In Section~\ref{pheno}, we briefly introduce the concept of supersonic turbulent energy cascade and, using dimensional arguments, 
show that scaling exponents in the linewidth--size and mass--size relations are algebraically coupled. We then derive a 
secondary relation between column density and size in several different ways and demonstrate consistency with 
observations, numerical models and theory. 
We argue that the turbulent interpretation of Larson's laws implies that {\em none of the three laws is fundamental}, but
that if one is known the other two follow directly from the scale invariance of the turbulent energy density transfer rate.
Section~\ref{gra} deals with the effects of self-gravity on small scales in star-forming clouds, and discusses the 
origin of the observed mass--size relation and the mass function for prestellar cores. 
Finally, in Section~\ref{fin} we formulate our conclusions and emphasize the statistical nature of the observed 
scaling laws.

\begin{table*}
\begin{minipage}{173mm}
    \caption{Simulation parameters.}
\begin{tabular}{l c c c c c c c c c c c l}
\hline
\hline
Model & $L$ & $N_{\rm root}$ & AMR & $\rho_0/m_{\rm H}$ & $u_{\rm rms}$ & $G$ & $B_0$ & EOS & $\chi_{\rm comp}$ & Section & Figure & Reference\\
& (pc) &&& (cm$^{-3}$) & (km s$^{-1}$) & & ($\mu$G) &&&&&\\
&(1)&(2)&(3)&(4)&(5)&(6)&(7)&(8)&(9)&(10)&(11)&(12)\\
\hline
HD1 &     1 & $2048^3$ & --- & 1 & 6 & 0 & --- & IT &  0   & \ref{im}, \ref{ms} &  \ref{s1}, \ref{dm} & \citet{kritsuk...09}\\
HD2 &     1 & $1024^3$ & --- & 1 & 6 & 0 & --- & IT & 0.4  & \ref{im}, \ref{ms} &  \ref{inter} & \citet{kritsuk...07a}\\
HD3 &     5 & $512^3$   & $5\times4$ & $10^3$ & 6 & 1 & --- & IT &  0.4  & \ref{img}, \ref{gra} &  \ref{s12}, \ref{fil} & \citet{kritsuk..11a}\\
MD1 & 200 & $512^3$   & --- &  5 & 16 & 0 & 0.95 & MP &  0  & \ref{mp} &  \ref{map}, \ref{s13} & \citet{kritsuk..11}\\
MD2 & 200 & $512^3$   & --- &  5 & 16 & 0 & 3.02 & MP &  0  & \ref{mp} &  \ref{map}, \ref{s13} & \citet{kritsuk..11}\\
MD3 & 200 & $512^3$   & --- &  5 & 16 & 0 & 9.54 & MP &  0  & \ref{mp} &  \ref{map}, \ref{s13} & \citet{kritsuk..11}\\
\label{params}
\end{tabular}
{\em Notes.} For each model, the following information is provided: (1) domain size in parsecs, dimensionless for scale-free 
models HD1 and HD2; (2) uniform (root) grid resolution; (3) AMR settings: number of levels $\times$  refinement factor;  
(4) mean number density in cm$^{-3}$, dimensionless for the scale-free models HD1 and HD2; (5) rms velocity in km s$^{-1}$, 
replaced by the rms Mach  number for isothermal models; (6) boolean gravity switch: 1 if included, 0 -- otherwise;
(7) the mean magnetic field strength;  (8) equation of state: isothermal (IT) or multiphase (MP);
(9) $\chi_{\rm comp}$ is the ratio of dilatational-to-total power in the external large-scale acceleration driving the turbulence; 
(10) Sections, where the results of the particular simulation are
presented and discussed; (11) Figures, where the data from the simulation are presented. 
\end{minipage}
\end{table*}

\section{Simulations}\label{sim}
In our analysis, we rely on data from a suite of six ISM simulations that include the effects of self-gravity, turbulence, magnetic 
fields and multiphase thermodynamics. The simulations use cubic computational domains with periodic boundary conditions.
All models use large-scale forcing \citep{maclow99} to mimic the energy flux incoming from scales larger than the box size. 
In all cases, the power in the driving acceleration field is distributed approximately isotropically and 
uniformly in the wavenumber interval $k/k_{\rm min}\in[1, 2]$, where $k_{\rm min}=2\pi/L$ and $L$ is the domain size.
For convenience, we further divide the models in two classes, hydrodynamic (HD) and magnetohydrodynamic (MD); 
Table~\ref{params} summarizes model parameters and physics included in the simulations.

The set of three HD simulations were run with the {\sc enzo} code\footnote{\tt http://enzo-project.org} \citep{enzo13},
which uses the third order accurate piecewise parabolic method \citep[PPM;][]{colella.84}. The MD simulations employed
an MHD extension of the method developed by \citet{ustyugov...09}. Model HD1 and all MD models used purely 
solenoidal driving, while models HD2 and HD3 were driven with a mixture of solenoidal and compressive motions. 
The inertial range ratio of compressive-to-solenoidal motions, and its dependence on Mach number and magnetic 
fields, is discussed in \citet{kritsuk...10}. All HD models assume an isothermal equation of state (EOS). As models
HD1 and HD2 do not include self-gravity, they only depend on the sonic Mach number, details of forcing aside. 
In addition, three dimensional parameters: (i) the box size, $L$; (ii) the mean density, $\rho_0$; and (iii) the temperature, 
$T$ (or sound speed, $c_{\rm s}$), fully determine the physical model. Typical values corresponding to MC conditions: 
$L=1-5$~pc, $\rho_0/(\mu m_{\rm H})=10^{2-3}$~cm$^{-3}$, and $c_{\rm s}=0.2$~cm~s$^{-1}$ for $T=10$~K.
The rms Mach number was chosen to be approximately 6, which is a good choice for the adopted range of $L$.
Models with substantially larger physical domain sizes cannot take advantage of a simple isothermal EOS. 

The HD3 simulation with self-gravity was earlier presented in \citet*{padoan...05} and \citet{kritsuk..11a}.
Model HD3 was initiated on a $512^3$ grid with a uniform density and a random initial large-scale velocity field. 
The implementation of driving in the {\sc enzo} code followed \citet{maclow99}, wherein small static velocity perturbations 
are added 
every time step such that kinetic energy input rate is constant. Once turbulence has had enough time to develop and 
statistically stationary conditions were reached, we switched gravity and adaptive mesh refinement (AMR) on and 
halted the driving. Note that
the kinetic energy decay in the short duration of the collapse phase is insignificant. Five levels of AMR were used 
(with a refinement factor of 4) conditioned on the Jeans length following the \citet{truelove.....97} numerical stability 
criterion. The model thus covered a hierarchy of scales from 5~pc (the box size) down to 2 au with an effective grid
resolution of $(5\times10^5)^3$.

The multiphase models MD1--3 represent a set of simple periodic box models, which ignore gas 
stratification and differential rotation in the disc and employ an artificial large-scale 
solenoidal force to mimic the kinetic energy injection from various galactic sources. 
This naturally leads to an upper bound on the box size, $L$, which determines our choice of 
$L=200$~pc. The MD models are fully defined by the following three parameters (all three would ultimately depend on $L$): (i)
the mean gas density in the box, $n_0$; (ii) the rms velocity, $u_{\rm rms,0}$; and (iii) the mean magnetic 
field strength, $B_0$, see Table~\ref{params} for numeric values and grid resolutions. 
The models also assume a volumetric heating source due to the far-ultraviolet (FUV) background radiation. 
This FUV radiation from OB associations of quickly evolving massive stars that form in MCs
is the main source of energy input for the neutral gas phases and this source is in turn balanced by radiative 
cooling \citep{wolfire....95}.
The MD models were initiated with a uniform gas distribution with an addition of small random isobaric 
density perturbations that triggers a phase transition in the thermally bi-stable gas that quickly turns 
$\sim25-65$ per cent of the gas mass into the thermally stable cold phase (CNM with temperature below $T=184$~K), 
while the rest of the mass is shared between the unstable and stable warm gas (WNM). The CNM and WNM each 
contain roughly $\sim 50$ per cent of the total H{\sc i} mass in agreement with observations \citep{heiles.05}. 
We then turn on the forcing and after a few large-eddy turnover times ($\sim10$~Myr) the simulations approach 
a statistical steady state. If we replace this two-stage initiation process with a one-stage 
procedure by turning the driving on at $t=0$, the properties of the steady state remain unchanged.
The rms magnetic field is amplified by the forcing and saturates when the relaxation in the system results 
in a steady state. The level of saturation depends on $B_0$ and on the rate of kinetic 
energy injection by the large-scale force, which is in turn determined by $u_{\rm rms}$ 
and $n_0$. This level can be easily controlled with the model parameters. In the 
saturated regime, models MD2 and MD3 tend to establish the kinetic/magnetic energy equipartition 
while the saturation level of magnetic energy in model MD1 is a factor of $\sim3$ lower than 
the equipartition level. The mean thermal energy also gets a slight boost due to the forcing, 
but remains subdominant in all three models. More detail on the MD models will follow in \S~\ref{mp}.

\section{The linewidth--size relation}\label{lws}
Recently, new methods of statistical analysis have been developed that provide diagnostics less ambiguous
than the original linewidth--size scaling based on the estimates of velocity dispersion and cloud size or on 
two-point statistics of the emission line centroid velocity fluctuations \citep{miesch.94}. One
example is the multivariate statistical technique of principal component analysis \citep[PCA;][]{heyer.97}, 
allowing one to extract the statistics of turbulent interstellar velocity fluctuations in a form that can be directly 
compared to the velocity statistics routinely 
obtained from numerical simulations. Using the PCA technique, \citet{heyer.04} found that 
the scaling of velocity structure functions (SFs) of 27 GMCs is close to invariant, 
\begin{equation}
S_1(u, \ell)\equiv\langle\left|\delta{}u_{\ell}\right|\rangle=u_0\ell^{0.56\pm0.02},
\label{hb04}
\end{equation}
for structures of size $\ell\in[0.03,50]$~pc \citep[see also][]{brunt03a,brunt03b}. 
Here $\delta u_{\ell}=u(r)-u(r+\ell)$ is the velocity difference between two
points in a three-dimensional volume separated by a lag $\ell$. Note that the lengths $\ell$ entering this relation
are the characteristic scales of the PCA eigenmodes; therefore, they may differ from the cloud sizes defined in
other ways \citep[e.g.][]{mckee.07}. See also \citet{romanduval+11} for a detailed discussion of
the relation between the PCA-determined SFs and `standard' first-order SFs obtained from simulations.

\citet{heyer...09} later used observations of a lower opacity tracer, $^{13}$CO, in 162 MCs 
with improved angular and spectral resolution to reveal weak systematic variations of the scaling 
coefficient, $u_0$, in (\ref{hb04}) with $\ell$ and $\Sigma$. Motivated by the concept
of clouds in virial equilibrium, they introduced a new scaling coefficient
$u_0^{\prime}\equiv\langle\left|\delta{}u_{\ell}\right|\rangle\ell^{-1/2}\propto\Sigma_{\ell}^{0.5}$.
This correlation, if indeed observed with a high level of confidence, would indicate a departure from `universality' for the velocity SF
scaling~(\ref{hb04}) found earlier by \citet{heyer.04} and compliance with the virial equilibrium condition~(\ref{vir})
advocated by SRBY87. 

Various scenarios employed in the interpretation of observations create a confusing picture of the structure formation 
in star-forming clouds carefully documented in a number of reviews 
\citep[e.g.][]{blitz93,elmegreen.04,maclow.04,ballesteros...07,mckee.07,hennebelle.12}. 
How can the apparent conspiracy between turbulence and gravity in MCs be understood and resolved? 
Analysis of data from numerical experiments will help us to shed some light on the nature of the confusion
in three subsections that follow.

\begin{figure}
\includegraphics[width=84mm]{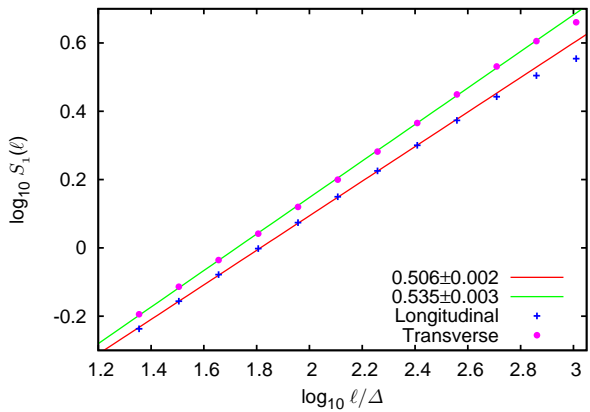}
\caption{Scaling of the first-order transverse (green) and longitudinal (red) velocity SFs 
in model HD1 \citep[compare with HD2, fig.~10 in][]{kritsuk...07a}.
$\Delta$ is the grid spacing. Numbers in the legend represent power-law indices for the linear 
least-squares representation of the data in  the range $\log_{10}(\ell/\Delta)\in[1.2,2.7]$ and formal uncertainty of the fits. 
}
\label{s1}
\end{figure}

\begin{figure*}
\begin{tabular}{cc}
\includegraphics[width=82mm]{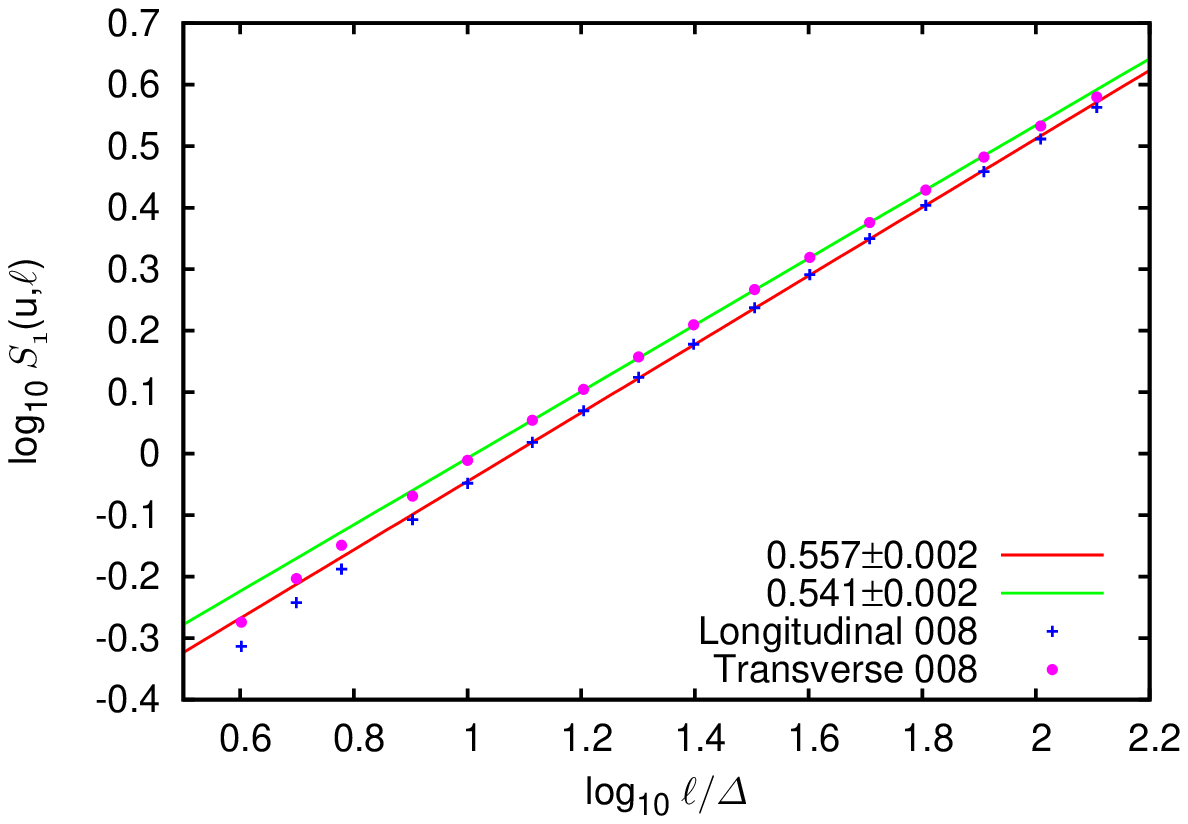} & \includegraphics[width=82mm]{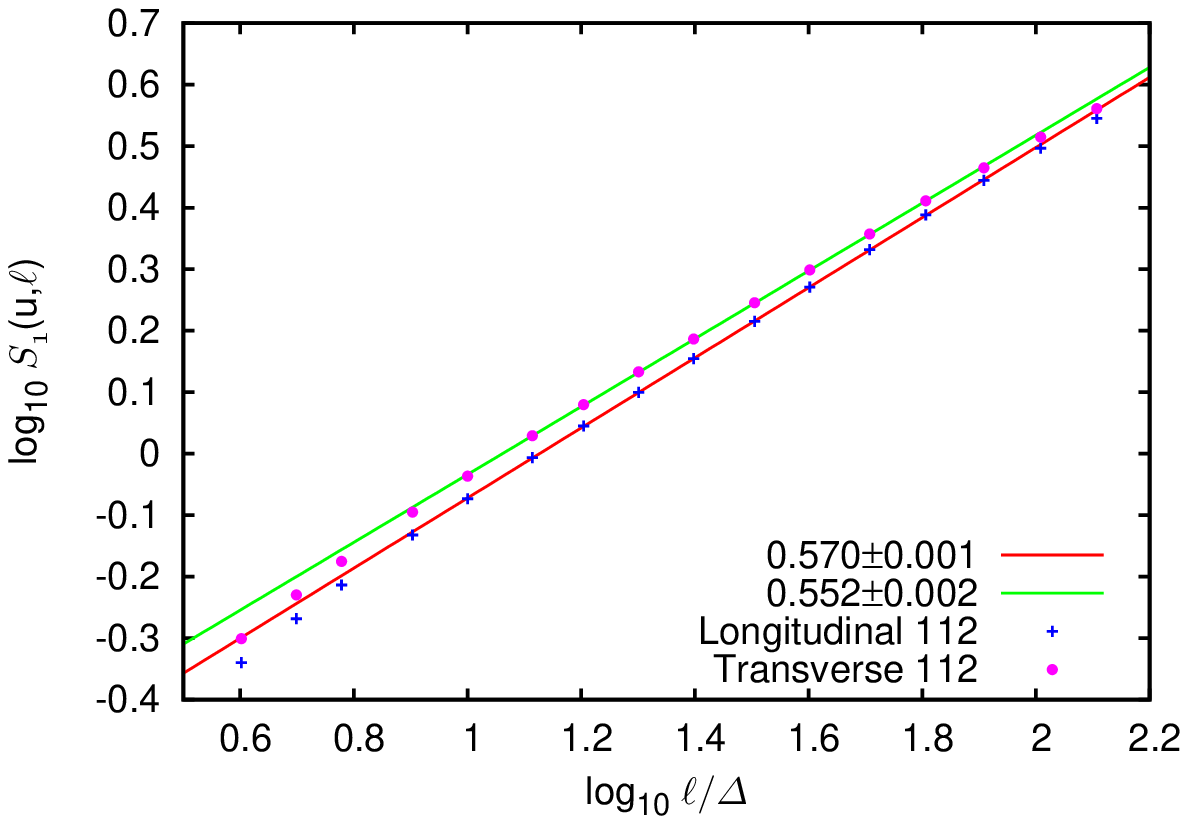}
\end{tabular}
\caption{As Fig.~\ref{s1}, but for an AMR simulation with self-gravity, model HD3.
The two plots show SF scaling based on the $512^3$ root grid information for two flow snapshots. Left-hand panel shows
the scaling in fully developed turbulence at $t=0$ (no self-gravity). Right-hand panel shows the same scaling plot, but for
the last flow snapshot at $t=0.43\,t_{\rm ff}$ (with self-gravity), when a number of collapsing cores are present in the computational domain. 
Each data point represents an average over $\sim8\times10^8$ velocity difference measurements. Numbers in the legend 
represent the power-law indices, $\zeta_1$, for the linear least-squares representation of the data points in the range $\ell/\Delta\in[10,100]$ 
and formal uncertainty of the fits. The root grid resolution in this model is $\Delta\simeq0.01$~pc; the range of scales shown
0.3--1.5~pc.}
\label{s12}
\end{figure*}

\subsection{Isothermal models}\label{im}
Numerical simulations of isothermal supersonic turbulence without self-gravity render inertial range scaling 
exponents for the first-order velocity SFs that are similar to those measured by \citet{heyer.04}. Model HD2
gives $S_1(u,\ell)\propto\ell^{\zeta_1}$ with $\zeta_{1,\parallel}=0.53\pm0.02$ 
and $\zeta_{1,\perp}0.55\pm0.04$ for longitudinal and transverse SFs, respectively.
The rms sonic Mach number used in these simulations ($M_{\rm s}=6$) corresponds to 
MC size $\ell\approx2$~pc, i.e. right in the middle of the observed scaling range. The simulation results indicate that 
velocity scaling in supersonic regimes strongly deviates from Kolmogorov's predictions for fluid turbulence. This 
removes one of the SRBY87 arguments against Larson's hypothesis of a turbulent origin for the linewidth--size 
relation. Indeed, the scaling exponent of $0.50\pm0.05$ measured in SRBY87 for whole clouds and a more
recent and accurate measurement $0.56\pm0.02$ by \citet{heyer.04} that includes cloud substructure both fall 
right within the range of expected values for supersonic isothermal turbulence at relevant Mach numbers.
Note, however, that the distinction between scaling `inside clouds' and `between clouds' often used in 
observational literature is specious if clouds are not isolated entities but rather part of a turbulence field
(they are being constantly buffeted by their surroundings) on scales of interest even though they might look as isolated in projection 
\citep[e.g.][]{henriksen.84}. Also note that the scaling exponent measured in SRBY87 is based on the
velocity dispersion, i.e. it is more closely related to the exponent of the second-order SF of velocity, $\zeta_2/2$,
and should be compared to $\zeta_{2,\parallel}/2\approx0.48$ and $\zeta_{2,\perp}/2\approx0.49$ obtained in model HD2.

Numerical simulations indicate that the power-law index $\zeta_1$ is not universal, 
as it depends on the Mach number and also on the way the turbulence is forced. At higher Mach numbers, the slope is
expected to be somewhat steeper, while a reduction of compressive component in the forcing would make it somewhat
shallower. \citet{schmidt....09} obtained $\zeta_{1,\perp}=0.54\pm0.01$ for purely compressive forcing at $M_{\rm s}\simeq2.3$ 
in a simulation with grid resolution of $768^3$. If the force is solenoidal, velocity SFs tend to show slightly smaller exponents due to 
an effective reduction in compressive motions at large scales. Naturally, the effect is more pronounced in SFs of 
longitudinal velocities, which track compressive motions more closely. Fig.~\ref{s1} illustrates this point with data 
from model HD1, where 
$\zeta_{1,\parallel}\simeq0.51$ and $\zeta_{1,\perp}\simeq0.54$.

The significance of the scaling dependence on the forcing found in numerical simulations should not be overestimated, 
though, due to their typically very limited dynamic range. It is not a priori clear whether the integral scale of the turbulence 
has to be the same in simulations with the same grid resolution that employ widely different forcing. In supersonic 
turbulence at very high Reynolds numbers, the solenoidal and dilatational modes are locally strongly coupled due to 
non-linear mode interactions in spectral space \citep{kritsuk...07a}. This locks the dilatational-to-solenoidal ratio at 
the geometrically motivated natural level of $\sim1/2$ \citep{nordlund.03}, which appears to be a universal asymptotic 
limit supported by turbulence decay simulations \citep{kritsuk...10}. 
Whether or not currently available models with compressive forcing allow enough room in spectral space to attain 
this level within the inertial range is a matter of debate. These concerns are supported by a resolution study
that shows substantially larger variation of relative exponents of first-order SFs when grid resolution increases from 
$512^3$ to $1024^3$ in models with compressive forcing ($\approx5$ per cent) compared to solenoidal ones 
\citep[$\approx2$ per cent, see table~II in][]{schmidt..08}. If the effective Reynolds numbers of compressively driven 
supersonic turbulence models are substantially smaller than those with natural forcing, then a meaningful comparison 
of inertial range scaling should involve compressive models with much higher grid resolutions to provide comparable 
scale separation in the inertial range. This would likely make such a comparison computationally expensive. 
Existing computational models with compressive forcing can also be verified using analytical relations for inertial range
scaling in supersonic turbulence \citep*{falkovich..10,galtier.11,wagner...12,aluie13,kritsuk..13}.
Since the Reynolds numbers in the ISM are quite large, directly comparing isothermal simulations with exotic forcing 
at grid resolutions $\lsim1024^3$ with observations is difficult due to restrictions imposed by the adopted EOS 
\citep*[cf.][]{federrath..09,federrath....10,romanduval+11}. 
In this respect, $\zeta_1\simeq0.54$ obtained in model HD2 with  forcing close to natural 
better represents the universal trends and thus bears more weight in comparison with relevant observations.

As we have seen, state-of-the-art observations leave substantial freedom of interpretation. In these circumstances,
numerical experiments can provide critical tests to verify or invalidate the proposed scenarios. One 
way to do this is to check the scaling of the velocity SFs in numerical simulations that include self-gravity and see 
if there is any difference. 

\subsection{Isothermal models with self-gravity}\label{img}
Here, we use  a very high dynamic range  simulation of isothermal self-gravitating turbulence with adaptive 
mesh refinement and effective linear resolution of $5\times10^5$ \citep{padoan...05,kritsuk..11a}. 
Our model HD3 is defined by the periodic domain size, $L=5$~pc, rms Mach number, $M_{\rm s}=6$, 
virial parameter, $\alpha_{\rm vir}\equiv5\sigma^2_{u}R/(GM)\approx0.25$ \citep{bertoldi.92}, free-fall time, 
$t_{\rm ff}\equiv(3\pi/32G\rho)^{1/2}\approx1.6$~{\rm Myr}, and dynamical time, 
$t_{\rm dyn}\equiv L/(2M_{\rm s}c_{\rm s})\approx2.3$~{\rm Myr}. The simulation was initialized as 
a uniform grid turbulence model by stirring the gas in the computational domain for $4.8t_{\rm dyn}$ 
with a large-scale random force that includes 40 per cent dilatational and 60 per cent solenoidal power.
After the initial stirring
period, which ended at $t=0$ with a fully developed statistically stationary supersonic turbulence, the forcing was 
turned off and the model was further evolved with AMR and self-gravity for about $0.29t_{\rm dyn}\simeq0.43t_{\rm ff}$.
We compare scaling of the first order velocity SFs computed for $t=0$ and $t=0.43t_{\rm ff}$ flow snapshots.
Fig.~\ref{s12} illustrates the result: a 1.8 per cent difference in scaling exponents 
between the non-self-gravitating ($\zeta_1\simeq0.55$, left-hand panel) and self-gravitating ($\zeta_1\simeq0.56$, 
right-hand panel) snapshots at the root grid resolution of $512^3$. The difference is small compared to the usual statistical 
snapshot-to-snapshot variation in turbulence simulations at this resolution. 

We clearly see  the lack of statistically significant 
imprints of self-gravity on the velocity scaling, even though by the end of the simulation the peak density has 
grown by more than eight decades and the density probability density function (PDF) has developed a strong power-law tail at the high end, 
while the initial lognormal low end of the distribution has shifted to slightly lower densities \citep{kritsuk..11a}.
Similar results were obtained by \citet{collins......12}, who independently demonstrated that in driven MHD turbulence 
simulations with self-gravity and AMR the velocity power spectra do not show any signs of ongoing core 
formation, while the density and column density statistics bear a strong gravitational signature on all scales. 
The absence of a significant signature of gravity in the velocity statistics can be readily
understood as the local velocity gains due to gravitational acceleration are simply insufficient to
be seen on the background of supersonic turbulent fluctuations supported by the large-scale 
kinetic energy injection during the stirring period. 

\begin{figure}
\includegraphics[width=84mm]{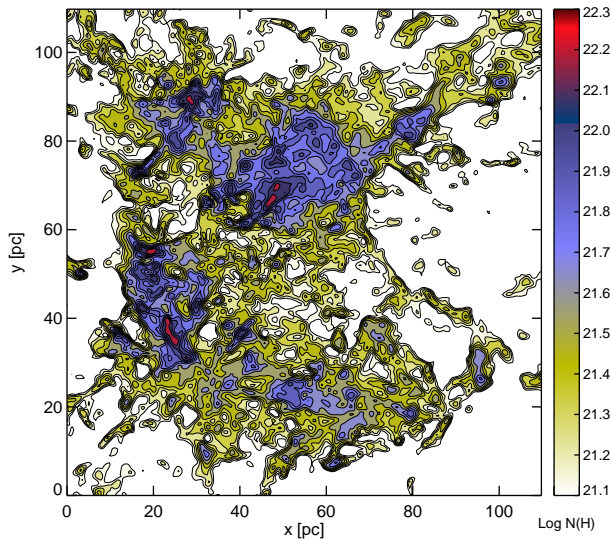}
\caption{A synthetic contour map of projected density distribution for the cold gas ($T<184$~K) in a subvolume of the 
(200~pc)$^3$ computational domain in the multiphase MHD turbulence model MD2 representing 
a simulated equivalent of an MC complex. The original grid 
resolution of the snapshot, $\Delta=0.39$~pc, has been coarsened by a factor of $\sim3$ to match the $0.5^{\circ}$
angular resolution of the historical $^{12}$CO map of MCs in Perseus, Taurus, and Auriga published 25 years ago by 
\citet{ungerechts.87}. The morphology of contours closely resembles the observations. Colour bar indicates the logarithm 
of column density of cold material.}
\label{map}
\end{figure}

We believe it is clear enough that as far as the inner velocity structure of MCs is concerned, it is obviously shaped 
by the turbulence. If there is a contribution from gravity, it is relatively small. The larger scale velocity picture can 
nevertheless bear more complexity, in 
particular at scales approaching the `full 3D system size' or the effective energy injection scale of ISM turbulence, which 
is believed to be close to the (cold) gaseous disc scaleheight. There, the formation of coherent, possibly gravitationally 
bound structures in an essentially two-dimensional setting of a differentially rotating disc would perhaps dominate over 
the seeds of direct three-dimensional turbulent energy cascade entering the inertial range.

Isothermal simulations HD1, HD2, and HD3 were designed to model the structure of MCs
on scales $\lsim5$~pc powered by the cascade coming from larger scales $\sim100$~pc, where this energy is 
injected by stellar feedback, galactic shear, gas infall on to the disc, etc. We have shown that isothermal models 
successfully reproduce the power index in the observed linewidth--size relation. We have also shown that self-gravity 
of the turbulent molecular gas, while actively operating on scales $\lsim1$~pc, does not produce a measurable 
signal in the linewidth--size relation. This suggests that the turbulence alone might indeed be responsible for the first 
Larson law, as observed within MCs, with a caveat that the coefficient in the relation cannot be predicted by 
the scale-free isothermal simulations by their design. Since an isothermal approximation we used so far cannot 
be justified for models of larger scale MC structure, we designed a numerical experiment 
that would properly model the ISM thermodynamics and allow us to define a reasonable proxy for the cold 
molecular gas traced by CO observations. 

\begin{figure*}
\begin{tabular}{cc}
\includegraphics[width=82mm]{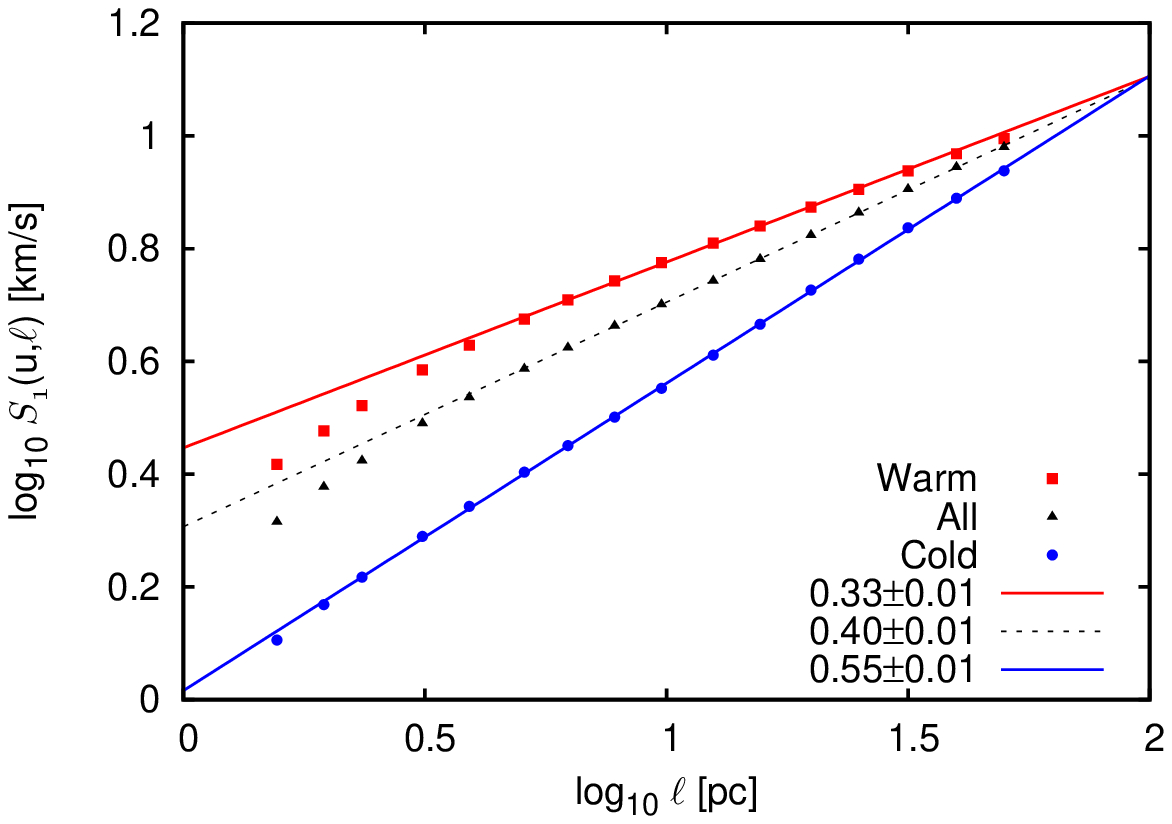}&
\includegraphics[width=82mm]{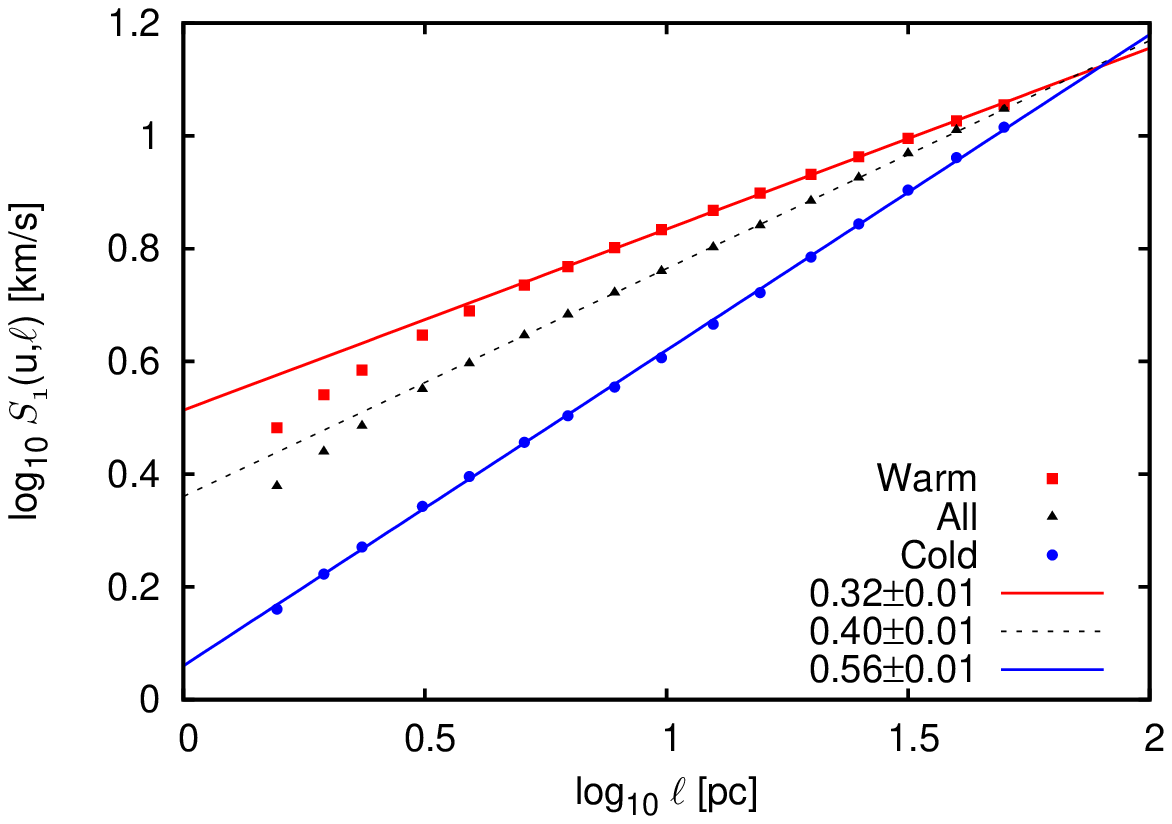}
\end{tabular}
\caption{First-order longitudinal (left) and transverse (right) velocity SFs based on the $512^3$ multiphase MHD turbulence 
simulation. Triangles represent
SFs for the whole domain; squares represent two-point correlations conditioned for the warm stable phase with temperatures
$T>5250$~K that fills $\sim23$ per cent of the volume; circles represent SFs conditioned for the cold phase with $T<184$~K that 
occupies $\sim7$ per cent by volume and contains $\sim50$ per cent of the mass.
Each triangle represents an average over $\sim6\times10^{10}$ velocity difference measurements in $\sim70$ flow snapshots covering
$\sim4\,t_{\rm dyn}$ or $\sim25$~Myr of statistically steady evolution. Numbers in the legend 
represent the power-law indices, $\zeta_1$, for the linear least-squares representation of the data points 
and the formal uncertainty of the fits.}
\label{s13}
\end{figure*}

\subsection{Multiphase models\label{mp}}
A suite of such numerical simulations, which also include an approximation for the cooling and heating 
functions \citep{wolfire....95} and magnetic field, but ignore self-gravity and local 
feedback from star formation, is described in section~3 of \citet*{kritsuk..11}. Here, we shall mostly discuss 
model MD2, a moderately magnetized case, with a domain size $L=200$~pc, mean hydrogen number density 
$n=5$~cm$^{-3}$, uniform magnetic field $B_0=3$~$\mu$G and a solenoidal random force acting on scales
between 100 and 200~pc. The force is normalized to match the width of the mass-weighted PDF of thermal pressure
in the diffuse cold medium of the Milky Way recovered from the {\em Hubble Space Telescope} 
observations \citep{jenkins.11,kritsuk..11}. The
model was evolved for $\sim60$~Myr (the dynamical time for model MD2 
$t_{\rm dyn}=L/2u_{\rm rms}\simeq6.1$~Myr) from 
uniform initial conditions and the final stretch of $\sim25$~Myr represents an 
established statistical steady state with an rms magnetic field strength $B_{\rm rms}\simeq11.7$~$\mu$G and 
with $\sim51$ per cent of the total gas mass $M_{\rm tot}=9.2\times10^5$~M$_{\odot}$ residing in the thermally stable 
cold phase at temperatures $T\lsim184$~K. 

Fig.~\ref{map} shows a synthetic column density map of what observationally would be classified as an MC
complex, morphologically resembling clouds in Perseus, Taurus, and Auriga observed in CO \citep{ungerechts.87}.
The map covers the cold gas in a projected area of $\sim30$ per cent of a full 2D flow snapshot. With a set of $\sim70$ 
snapshots available for this model, one can easily track the formation, evolution, and breakdown of structure in the 
MC complex over a few $10^7$~years. 
While in projection the mapped MCs may look like quasi-uniform clouds with well-defined boundaries, the dense cold 
clumps of gas in the simulation box actually fill only a small fraction of their volume ($\sim7$ per cent overall), similar to what 
observations suggest \citep[e.g.][]{blitz87}. In three dimensions, the clouds represent a collection of mostly disjoint, 
under-resolved structures that randomly overlap in projection. As these structures move supersonically, the physical 
identity (in the Lagrangian sense) of the `clouds' we see in projection is actively evolving. This makes simplified scenarios 
often discussed in the literature, which imply isolated static clouds (e.g. a collision of two clouds or collapse of 
an isolated cloud) or converging flows taken out of the larger-scale turbulence context very unrealistic. 
The so-called interclump medium \citep[ICM; e.g.][]{hennebelle.12} 
is mostly composed of thermally unstable gas at intermediate temperatures \citep{field65,hunter70,heiles01,kritsuk.02}.
The ICM comprises $\sim70$ per cent of the whole volume of the computational domain and $\sim23$ per cent is filled with the 
stable warm phase at $T\gsim5250$~K. 
Phase transformations are quite common in such multiphase environments with individual fluid elements actively cycling 
through various thermal states in response to shocks and patches of supersonic velocity shear. 

Turbulence models of this type bear substantially more complexity since different thermal phases represent different 
physical regimes -- from supersonic and super-Alfv\'enic in the cold phase, including the molecular gas, 
($M_{\rm s}\simeq15.2$, $M_{\rm A}\simeq4.2$) to transonic and trans-Alfv\'enic in the warm phase ($M_{\rm s}\simeq1.7$, 
$M_{\rm A}\simeq1.2$). Active baroclinic vorticity creation fosters exchange between compressible kinetic and thermal
modes otherwise suppressed by kinetic helicity conservation enforced in isothermal turbulence \citep{kritsuk.02,elmegreen.04}.
This rich diversity ultimately determines velocity scaling and other statistical properties of multiphase turbulence \citep{kritsuk.04}. 

In order to check if model MD2 will reproduce the velocity scaling recovered from observations of tracers of the cold gas, 
such as $^{12}$CO or  $^{13}$CO, we can calculate velocity SFs conditioned on the cold phase only, i.e. all 
velocity differences entering the ensemble average will be measured for point pairs located in the cold gas only.
While a detailed knowledge of CO chemistry is needed to connect models and observations, this simplified approach
can be used as the first step from isothermal models to more sophisticated and costly numerical simulations with 
complex interstellar chemistry.

Fig.~\ref{s13} shows time-averaged first-order SFs of longitudinal (left) and transverse (right) velocity in the 
statistically stationary regime of model MD2. Black triangles indicate SFs for point pairs selected from the whole volume
following the same procedure as in Figs~\ref{s1} and \ref{s12}. The scaling exponent $\zeta_{1,\rm all}\simeq0.4$ is 
somewhat larger than the Kolmogorov value of $1/3$ as most of the volume is filled with mildly supersonic thermally 
unstable gas ($M_{\rm s}\simeq4.2$). Red triangles show SFs conditioned on the warm stable phase only. Since
the warm phase is essentially transonic, the velocity scaling is very close to Kolmogorov's,  
$\zeta_{1,\rm warm}\simeq0.33$. Blue circles indicate SFs conditioned on the cold, thermally stable, strongly supersonic phase.
The exponents, $\zeta_{1,\rm cold}\simeq0.56$, measured in the range $\ell\in[1.5, 50]$~pc are consistent with the 
results of CO observations we discussed earlier in this section. 

The slope of the correlation is only weakly sensitive 
to the magnetization level assumed in the model. Model MD3 has $B_0=9.5$~$\mu$G, $B_{\rm rms}\simeq16$~$\mu$G 
and $\zeta_{1,\rm cold}\simeq0.52$, while model MD1 has $B_0=1$~$\mu$G, $B_{\rm rms}\simeq8$~$\mu$G and 
$\zeta_{1,\rm cold}\simeq0.65$.

An interesting feature of the velocity scaling in the cold gas (circles in Fig.~\ref{s13}) -- which still needs to be understood -- is 
the extent of the power law into the range of scales where numerical dissipation usually plays a role and its effects 
are clearly seen in the unconditional SF scaling (triangles) and in the scaling for the warm gas (squares). 
Even though the number of point pairs in the cold phase is limited due to the small volume fraction of the phase, 
the scaling is still reasonably well defined statistically due to the large number of flow snapshots involved in the 
averaging.

Another important feature of the cold gas velocity scaling is the value of the coefficient $u_0(1$~pc$)\sim1$~km s$^{-1}$  
in models MD1, MD2 and MD3, independent of the magnetization level. This value roughly matches the observationally determined 
coefficient in the linewidth--size relation. As we mentioned above, the rms velocity in the box is controlled by the forcing 
normalization and by the mean density of the gas. These values were chosen to match the observed distribution of thermal 
pressure, which is only weakly sensitive to the magnetic effects. A much weaker or stronger forcing would change 
the rms velocity and the coefficient, but then the pressure distribution would not match the observations, see for instance 
model E in \citet{kritsuk..11}.

Finally, while near the energy injection scale $\sim100$~pc the velocity correlations for different thermal phases show the same amplitude,
there is a clear difference in the SF levels for the cold and warm gas in the inertial range. Due to the difference in SF slopes at different Mach 
number regimes probed, the warm gas shows progressively higher velocity differences than the cold gas at lower length-scales. 
For instance, at $\ell=10$~pc the ratio $\langle|\delta u_{\rm cold}|\rangle/\langle|\delta u_{\rm warm}|\rangle\sim0.6$. Our multiphase models 
only allow for a quite broad definition of the cold phase based on the temperature threshold determined by the thermal stability
criterion \citep{field65}. However one can imagine that observations of molecular gas tracers with different effective temperatures and densities 
would quite naturally provide different estimates for the velocity dispersion as a function of scale. This would in turn lead to different 
virial parameters probed by different traces with the net result of clouds observed using lower opacity traces to {\em appear} more 
strongly bound \citep[e.g.][]{koda...06}.

As a quick  summary of the results so far, we have seen that numerical simulations of interstellar turbulence can successfully
reproduce the observed linewidth--size relation including both the coefficient and the slope. Self-gravity of the gas does not 
seem to produce any significant modifications to the scaling relation over the range of scales simulated. 
Our isothermal models HD1--3 cover the range of scales from about 2~pc down to 0.3~pc, while the multiphase
models MD1--3 extend this range to $\sim50$~pc.

\section{The mass--size relation}\label{ms}
An alternative formulation of the original Larson's third law, $m\propto{}L^{1.9}$, implied a nested hierarchical
density structure in MCs. Such a concept was proposed by \citet{hoerner51} \citep[see also][]{weizsacker51} to 
describe an intricate statistical mixture of shock waves in highly compressible interstellar turbulence. 
He pictured density fluctuations as a hierarchy of interstellar clouds, analogous to eddies in incompressible 
turbulence. Observations indeed reveal a pervasive `fractal' structure in the interstellar gas that is usually interpreted 
as a signature of turbulence \citep{falgarone.91,elmegreen.96,lequeux05,romanduval....10,hennebelle.12}. 
Numerous attempts to measure the actual Hausdorff dimension of the MC structure either observationally 
or in simulations demonstrated that such measurements are notoriously difficult and bear substantial 
uncertainty, which depends on what measure is actually used. Here, we narrow down the discussion 
solely to the so-called mass dimension, i.e. the power index in the third Larson law. 

The most recent result for a sample of 580 MCs, which includes the SRBY87 clouds, shows a very 
tight correlation between cloud radii and masses, 
\begin{equation}
m(R)=(228\pm18{\rm M}_{\odot})R^{2.36\pm0.04},\label{rd10}
\end{equation}
for $R\in[0.2,50]$~pc  \citep{romanduval....10}. 
The power-law exponent in this relation, $d_{\rm m}\approx2.36$, corresponds to a `spongy' 
medium organized by turbulence into a multiscale pattern of clustered corrugated shock-compressed 
layers \citep{kritsuk..06}. Note, that $d_{\rm m}\approx2.36$ under a reasonable assumption of anisotropy 
implies $\Sigma\propto{}mL^{-2}\propto{}L^{0.36}$.
Thus, the observed mass--size correlation does not support the idea of a universal 
mass surface density of MCs. Meanwhile, a positive correlation of $\Sigma$ with $\ell$
removes theoretical objections against the third Larson law outlined in Section~1.

\begin{figure}
\includegraphics[width=84mm]{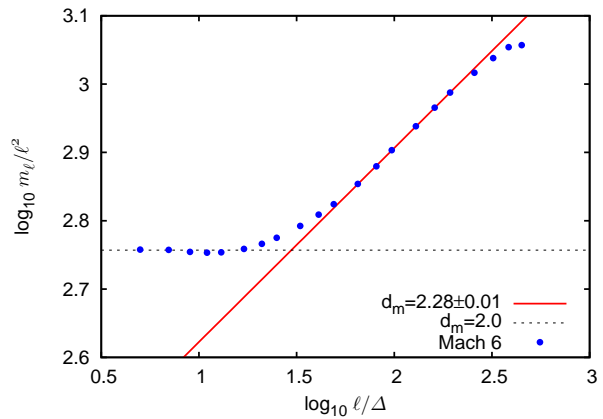}
\caption{Compensated scaling of the mass, $m_{\ell}$, with size $\ell$
in model HD1. Dissipation-scale structures are shocks with $d_{\rm m}\simeq2$. Inertial-range structures have
$d_{\rm m}\simeq2.3$.
}
\label{dm}
\end{figure}

Direct measurements of $d_{\rm m}$ in the simulations give the inertial subrange values 
in good agreement with observations: 
$d_{\rm m}=2.39\pm0.01$ (model HD2) and $2.28\pm0.01$ 
(model HD1; Fig.~\ref{dm}). Note that $\pm0.01$ represents the formal uncertainty of
the fit, not the actual uncertainty of $d_{\rm m}$, which is better characterized by the difference
between the two independent simulations and lies somewhere within $\pm(5-10)$ per cent of $d_{\rm m}$.
\citet{federrath..09} obtained $d_{\rm m}\simeq2.11$ and $2.03$ from two $1024^3$ simulations with purely
solenoidal and compressive forcing, respectively (see their fig.~7). Their result for solenoidal forcing is broadly 
consistent with our measurement for a larger $2048^3$ simulation within the uncertainty of the measurement.
However, their model with a purely compressive forcing rendered a lower mass dimension value of $2.03$,
while it is expected to be larger for this type of forcing. A detailed inspection of fig.~7 in
\citet{federrath..09} shows that the scaling range $d_{\rm m}(\ell)={\rm const.}$ for this model is 
quite short or nonexistent making this measurement very uncertain. This further indicates that concerns about 
insufficient scale separation in numerical simulations with purely compressive forcing (see Section~\ref{im}) are
valid.

An accurate measurement of $d_{\rm m}$ in numerical simulations of supersonic turbulence requires 
extensive statistical sampling since the density (unlike the velocity) experiences huge spatial and 
temporal fluctuations. Because of that, only the highest resolution simulations with sufficient scale 
separation can yield reliable estimates of $d_{\rm m}$. We have shown above that, like $\zeta_1$, 
the mass dimension also depends on forcing. However, as we discussed in Section~\ref{im}, the 
non-linear coupling of dilatational and solenoidal modes in compressible turbulence provides certain 
constraints on the artificial forcing employed in isothermal simulations \citep{kritsuk...10}. 
Therefore, the higher end values of mass dimension $d_{\rm m}\simeq2.4$ recovered from simulations 
of supersonic turbulence with forcing close to natural seem more realistic.

Observations of \citet{romanduval....10} represent perhaps the most extended sample 
of local clouds available in the literature \citep[for $^{12}$CO data compilations covering a wider Galactic 
volume and range of size scales $0.005\lsim L\lsim200$~pc, see][]{lequeux05,hennebelle.12}.

With $d_{\rm m}\simeq2.36$, the power-law index in the linewidth--size relation compatible 
with the virial equilibrium condition (\ref{vir}), $\zeta_{1,\rm vir}\equiv(d_{\rm m}-1)/2\simeq0.68$, 
is still reasonably close to the scaling exponent $\zeta_1\simeq0.56$ in equation~(\ref{hb04}), even if one 
assumes $u_0={\rm const.}$ (see Section~\ref{uni} below). Thus, we cannot exclude the possibility 
of virial equilibrium \citep[or kinetic/gravitational energy equipartition, see][]{ballesteros06} across 
the scale range of several decades based on the available observations alone. However, if one adopts
the virial interpretation of Larson's laws, the origin of the mass--size correlation observed over a wide range 
of scales remains unclear. The turbulent interpretation does not have this problem as the 
slopes of both linewidth--size and mass--size correlations are readily reproduced in simulations and 
can be predicted for the Kolmogorov-like cascade picture using dimensional arguments, see 
Section~\ref{pheno}. Moreover, the two slopes in the cascade picture are algebraically coupled and
this connection is supported observationally, as well as the inertial cascade picture itself 
\citep{hennebelle.12}.

\section{Interstellar turbulence and Larson's laws}\label{pheno}
In the absence of a simple conceptual theory of supersonic turbulence, many important questions related to the
structure and kinematics of MCs remained unanswered. One of them is the applicability of the universality concept to 
MC turbulence. The other related question is which of Larson's laws is fundamental? We briefly review the 
current theoretical understanding of universal properties of turbulence and address these two 
basic questions below using a phenomenological approach.

\subsection{Universality in incompressible turbulence}\label{uni}
Before we generalize our discussion to turbulence in compressible fluids, it is instructive to briefly review the 
notion of universality as it was first introduced for incompressible fluids. In turbulence research, universality 
implies independence of scaling on the particular mechanism by which the turbulence is generated \citep[e.g.][]{frisch95}. 
Assuming homogeneity, isotropy and finiteness of the energy dissipation rate in the limit of infinite Reynolds
numbers, $Re\equiv\ell_0u_0/\nu\rightarrow\infty$, \citet{kolmogorov41a,kolmogorov41b} derived the following exact relation
from the N-S equation
\begin{equation}
S_{3,\parallel}(u,\ell)\equiv\langle(\delta u_{\parallel,\ell})^3\rangle=-\frac{4}{5}\varepsilon\ell,
\label{45}
\end{equation}
known as the four-fifths law.  Here, $\delta u_{\parallel,\ell}$ is the longitudinal velocity increment
\begin{equation}
\delta u_{\parallel,\ell}=[\pmb{u}(\pmb{x}+\pmb{\ell})-\pmb{u}(\pmb{x})]\pmb{\cdot}\pmb{\ell}/\ell,
\end{equation}
$\varepsilon$ is the mean energy dissipation rate per unit mass and equation (\ref{45}) holds in the inertial 
subrange of scales $\eta\ll\ell\ll\ell_0$, where the direct influence of energy injection at 
$\ell_0\rightarrow\infty$ and dissipation at the Kolmogorov scale 
$\eta\equiv(\nu^3/\varepsilon)^{1/4}$ can be neglected. The four-fifths law is essentially a statement 
of conservation of energy in the inertial range of a turbulent fluid, while $S_{3,\parallel}(u,\ell)/\ell$ is
an indirect measure of the energy flux through spatial scale $\ell$, corresponding to the \citet{richardson22}
picture of the inertial energy cascade. Assuming further that the cascade proceeds in a self-similar way,
the K41 theory predicted a general scaling for $p$th order SFs
\begin{equation}
S_p(u,\ell)=C_p\varepsilon^{p/3}\ell^{p/3}, 
\label{41}
\end{equation}
where the constants $C_p$ are dimensionless. A particular case of equation (\ref{41}) with $p=2$ leads to the
famous Kolomogorov--Obukhov energy spectrum 
\begin{equation}
P(u,k)=C_{\rm K}\varepsilon^{2/3}k^{-5/3}, 
\label{53}
\end{equation}
where $k\equiv2\pi/\ell$ is the wavenumber and $C_{\rm K}$ is the Kolmogorov constant. The slope of the spectrum
$\beta=5/3$ is related to the slope of the second order SF $\zeta_2=2/3$ since one is the Fourier transform of
the other and therefore 
\begin{equation}
\beta=\zeta_2+1.
\label{ft}
\end{equation}
Likewise, if the self-similarity holds true, then equation (\ref{ft}) can be used to obtain the slope of the first-order SF as a function of
the power spectrum slope
\begin{equation}
\zeta_1=(\beta-1)/2.
\label{bur}
\end{equation}

The self-similarity hypothesis
implies that the statistics of velocity increments in equation (\ref{41}) are fully determined by the universal constants $C_p$ and 
$C_{\rm K}$, and depend only on the energy injection rate $\varepsilon>0$ and the lag $\ell$. This prediction, however, was
challenged by Landau in 1942 \citep[section~38, p.~140]{landau.87} and did not receive experimental support 
(for orders $p$ other than 3), eventually leading to the formulation of the refined similarity hypothesis (RSH)
\begin{equation}
S_p(u,\ell)=C_p\langle\varepsilon_{\ell}^{p/3}\rangle\ell^{p/3}\propto\ell^{\zeta_p}, 
\label{62}
\end{equation}
(where $\zeta_p=p/3+\tau_{p/3}$) to account for the inertial range intermittency \citep{kolmogorov62}. A hierarchical 
structure model based on the RSH and on the assumption of log-Poisson statistics of the dissipation field 
$\varepsilon_{\ell}$  \citep{she.94,dubrulle94} successfully predicted the values of anomalous scaling exponents $\zeta_p$
measured in large-scale direct numerical simulations (DNS). Recently obtained DNS data at large, but finite Reynolds 
numbers support the universal nature of anomalous exponents $\zeta_p$ in the spirit of K41, but including relatively 
small intermittency corrections $\tau_p\ne0$ for orders $p$ at least up to $p=8$, except $\tau_3=0$ \citep{gotoh02,ishihara..09}. 
The DNS results also show that the slope of the energy spectrum compensated with $k^{5/3}$ is slightly tilted, suggesting that
$P(u,k)\propto k^{-5/3+\phi}$ with $\phi=-0.1$ \citep{kaneda....03}. Whether the constants $C_p$ and $C_{\rm K}=1.6\pm0.1$ are universal 
or not still remains to be seen \citep[for references see][]{ishihara..09}.

\subsection{Universal properties of supersonic turbulence}\label{suni}
As far as compressible fluids are concerned, up until recently there were no exact relations similar to the four-fifths law (\ref{45})
available. However,  the growing body of theoretical work 
\citep{falkovich..10,galtier.11,aluie11,aluie..12,wagner...12,banerjee.13,aluie13} now supports the existence of a Kolmogorov-like 
inertial energy cascade in turbulent compressible fluids suggested earlier by phenomenological approaches 
\citep{henriksen91,fleck96,kritsuk...07a}. 

For the extreme case of supersonic turbulence in an isothermal fluid, which represents a simple model of particular interest 
for MCs, an analogue of the four-fifths law was obtained and verified with numerical simulations \citep{kritsuk..13}
\begin{equation}
\left\langle\delta(\rho \pmb{u})\cdot\delta\pmb{u}\,\delta u_{\parallel} +
[\delta(d\rho\pmb{ u})-\tilde{\delta}d\,\delta(\rho\pmb{ u})]\cdot\delta\pmb{ u}\right\rangle
\simeq - \frac{4}{3}\rho_0\varepsilon \ell,
\label{13}
\end{equation}
where $d\equiv\pmb{\nabla\cdot u}$ is the dilatation, $\tilde{\delta }d\equiv d(\pmb{x}+\pmb{\ell})+d(\pmb{x})$, 
$\varepsilon$ has the same meaning as in equation (\ref{45}), and $\rho_0$ is the mean density. This relation follows from
a more involved exact relation derived from the compressible N-S equation using an isothermal closure
\citep{galtier.11}. It further reduces to a primitive form of equation (\ref{45}) in the incompressible limit, 
\begin{equation}
\left\langle(\delta\pmb{u})^2\,\delta u_{\parallel}\right\rangle = - \frac{4}{3}\varepsilon \ell,
\label{131}
\end{equation}
when $\rho\equiv\rho_0$
and $d\equiv0$. 

At Mach 6 typical for the MC conditions, the first term on the l.h.s. of equation (\ref{13}) representing the
inertial flux of kinetic energy is negative and a factor of $\approx3$ larger in absolute value than the sum of the second 
and third terms describing an energy source associated with compression/dilatation. Numerical results show that 
both flux and source terms can be approximated as linear functions of the separation $\ell$ in the inertial range. 
Hence, equation (\ref{13}) reduces to
\begin{equation}
\left\langle\delta(\rho \pmb{u})\cdot\delta\pmb{u}\,\delta u_{\parallel}\right\rangle
\simeq - \frac{4}{3}\rho_0\varepsilon_{\rm eff} \ell,
\label{132}
\end{equation}
where the effective energy injection rate $\varepsilon_{\rm eff}$ includes the source contribution. Comparing equations
(\ref{131}) and (\ref{132}) one can see that strong compressibility in supersonic turbulence is ultimately accounted 
for by the presence of the momentum density increment in the l.h.s. of equation (\ref{132}) and also formally requires a 
compressible correction to the mean energy injection rate.

\begin{figure}
\includegraphics[width=84mm]{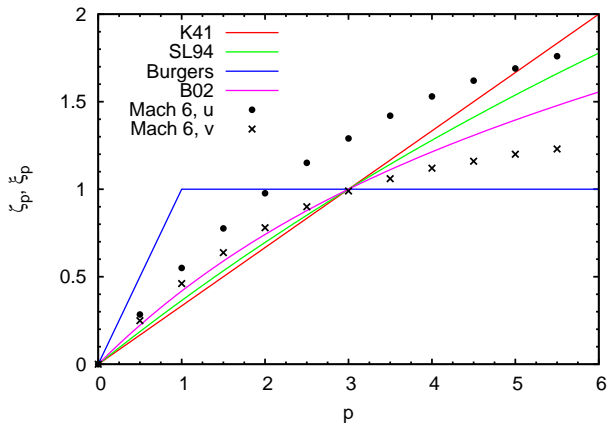}
\caption{Absolute scaling exponents for SFs of transverse velocity [$S_p(u,\ell)\propto\ell^{\zeta_p}$,
circles] and density-weighted velocity $v=\rho^{1/3}u$ [$S_p(v,\ell)\propto\ell^{\xi_p}$, crosses] for model HD2.
Solid lines show $\zeta_p$ predicted by the K41 theory (red), Burgers model (blue), and 
intermittency models due to \citet[][green]{she.94} and \citet[][magenta]{boldyrev02}. Note that the Burgers model 
predicts $\zeta_1=1$ rather than $\zeta_1=1/2$ \citep[cf.][]{mckee.07}.
}
\label{inter}
\end{figure}

A simplified formulation of equation (\ref{132}) based on dimensional analysis
\begin{equation}
S_3(v,\ell)\equiv\left\langle|\delta v|^3\right\rangle\propto\ell,
\label{133}
\end{equation}
where $\pmb{v}\equiv\rho^{1/3}\pmb{u}$, was introduced in \citet{kritsuk...07a,kritsuk...07b} and 
its linear scaling was further verified with numerical simulations in a wide range of Mach numbers 
\citep{kowal.07,schwarz...10,zrake.12}. Note that a `symmetric' density weighting in the third-order 
SF (\ref{133}) differs from the original form in the l.h.s. of equation (\ref{132}), which includes 
momentum  and velocity differences, and that the absolute value of the increment is taken in equation (\ref{133}). 
While the linear scaling is preserved in equation
(\ref{133}) for absolute values of both longitudinal and transverse differences of $\pmb{v}$, the 
extent of the scaling range is substantially shorter compared to that seen in equation (\ref{132}) for the 
same numerical model \citep{kritsuk..13}.

Finally, using (\ref{133}) and following an analogy to K41, one can also postulate self-similarity of the energy cascade in 
supersonic turbulence and predict a general scaling for the $p$th order SFs of the mass-weighted velocity
\begin{equation}
S_p(v,\ell)\equiv\left\langle|\delta v|^{p}\right\rangle\propto\ell^{p/3}
\label{134}
\end{equation}
and for the power spectrum
\begin{equation}
P(v,k)\propto k^{-5/3}. 
\label{533}
\end{equation}
While scaling relations (\ref{134}) and (\ref{533}) generally reduce to (\ref{41}) and (\ref{53}) in the incompressible limit,
it is hard to expect them to universally hold in compressible fluids for the following two reasons: 
First, as we have seen in \S~\ref{uni}, even in incompressible fluid flows, relations (\ref{41}) and 
(\ref{53}) are not strictly universal due to intermittency (except for $p=3$). 
Secondly, the transition from relationship (\ref{13}) to (\ref{133}) involves 
a chain of non-trivial assumptions that can potentially influence (or even eliminate) the scaling. 
A combination of these two factors can produce different results depending 
on the nature of the large-scale energy source for the turbulence.

Nevertheless, the power spectra of $\pmb{v}$ demonstrated Kolmogorov-like slopes $\beta\approx5/3$ in numerical 
simulations at moderately high Reynolds numbers, suggesting that intermittency corrections should be somewhat
larger than in the incompressible case, but still reasonably small at $p=2$ 
\citep{kritsuk...07a,kritsuk...07b,schmidt..08,federrath....10,price.10}. For instance, the `world's largest simulation' of
supersonic ($M=17$) turbulence with numerical resolution of $4096^3$ mesh points and a large-scale solenoidal forcing
gives $P(v,k)\propto k^{-5/3+\psi}$ with $\psi=-0.07$ \citep{federrath13}, which shows an intermittency correction $\psi$ 
similar to $\phi=-0.1$ seen in the incompressible DNS of comparable size \citep{kaneda....03}.

In a twin $4096^3$ simulation with purely compressive forcing limited to $k/k_{\rm max}\in[1,3]$, \citet{federrath13} 
finds a steep scaling range with $P(v,k)\propto k^{-2.1}$ at $k/k_{\rm max}\in[12, 30]$, which then flattens to $-5/3$ 
at $k/k_{\rm max}\in[40,50]$ shortly before entering the so-called `bottleneck' range of wavenumbers. It is not 
clear where the inertial range is in this case since detailed analysis of the key correlations in relation (\ref{13}) is lacking. 
Visual inspection of flow fields in this simulation indicates the presence of strong coherent structures associated with 
the large-scale energy injection, which can be responsible for the $-2.1$ scaling. It seems likely that in the extreme
case of purely compressive stirring, the mean energy injection is substantially less localized to the large scales
due to the nature of non-linear energy exchange between the dilatational and solenoidal modes \citep{moyal52},
resulting in a substantially shorter inertial range \citep[see also][]{kritsuk...10,wagner...12,aluie13,kritsuk..13}. 
A factor of $\approx4$ lower level of enstrophy in this model compared to the one with solenoidal forcing implies
a lower value of the effective Reynolds number, also symptomatic of a shorter inertial range.

\begin{figure*}
\begin{tabular}{cc}
\includegraphics[width=82mm]{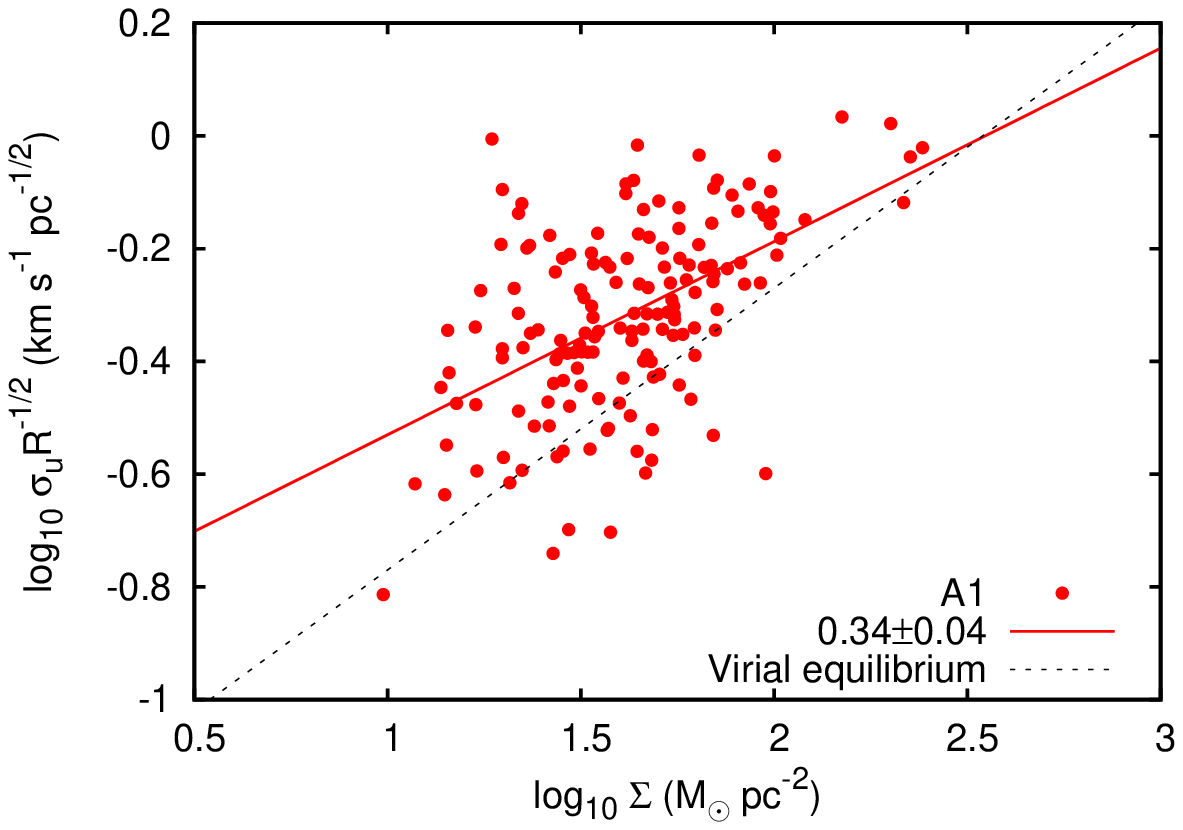}&
\includegraphics[width=82mm]{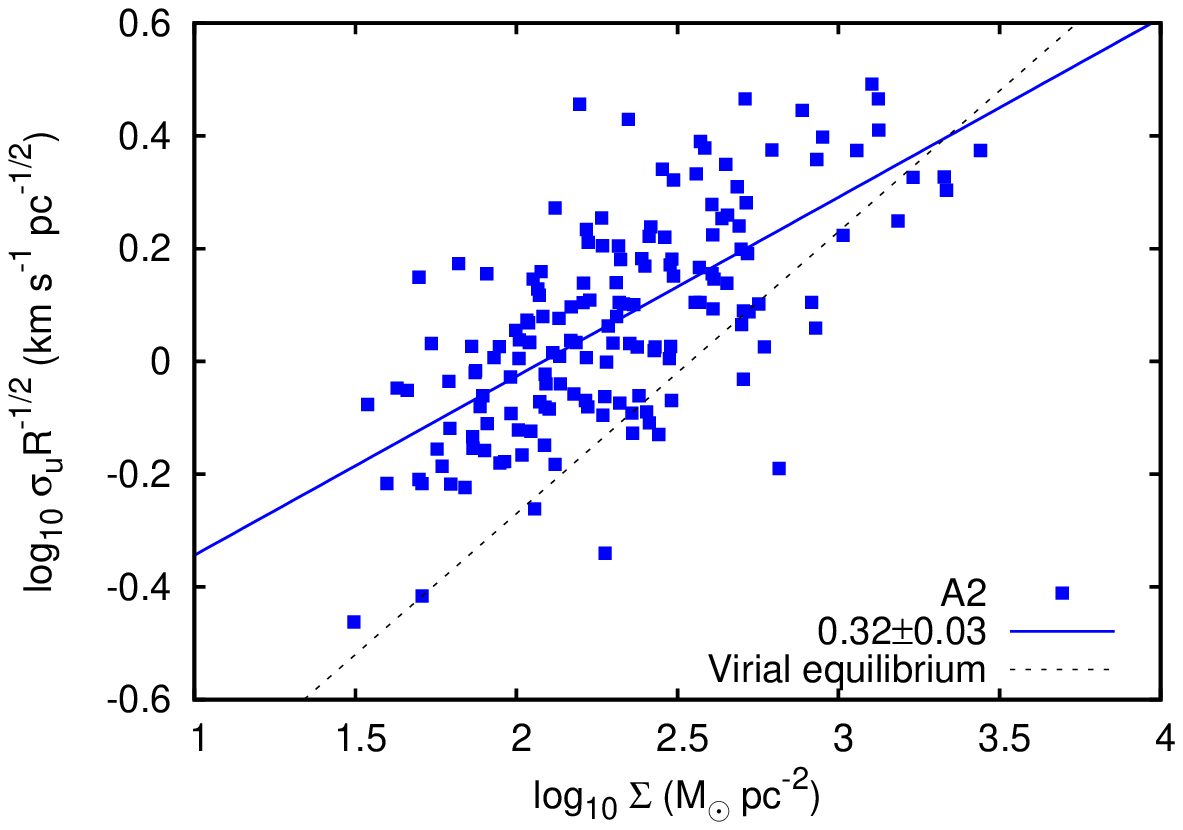}
\end{tabular}
\caption{Variation of the scaling coefficient $u_0^{\prime}=\sigma_u{}R^{-1/2}$ with mass surface density $\Sigma$
based on data from \citet{heyer...09}. Solid lines with slopes $0.34\pm0.04$ and $0.32\pm0.03$ show linear least-squares 
fits to A1 (left) and A2 (right) subsets from \citet{heyer...09}, respectively. Dashed line in both cases shows the 
correlation expected for clouds in virial equilibrium.
}
\vspace{0.02cm}
\label{sigma}
\end{figure*}

Fig.~\ref{inter} shows how the absolute scaling exponents of SFs of $\pmb{u}$ and $\pmb{v}$ vary with the order $p$
in model HD2 at $M_{\rm s}=6$. Filled circles show exponents $\zeta_p$ of the velocity SFs for $p\in[0.5,5.5]$.
In subsonic or transonic regimes these exponents would closely follow the K41 prediction $\zeta_p=p/3$, if isotropy 
and homogeneity were assumed and intermittency ignored. Starting from $M_{\rm s}\gsim3$, however, 
$\zeta_3$ shows a clear excess over unity that gradually increases with the Mach number, indicating a non-universal 
trend. For the density-weighted velocity ${\bmath v}$, the slope of the third-order SF, $\xi_3\approx1$, 
does not depend on the Mach number, while $\xi_p$ at $p\ne3$ demonstrate an anomalous scaling that is successfully
predicted by the hierarchical structure model of \citet{she.94}, although with parameters that differ from those for the
incompressible case and from those suggested by \citet{boldyrev02} for relative exponents $\zeta_p/\zeta_3$
\citep{kritsuk...07b,pan..09}.  Note that supersonic turbulence is more intermittent than incompressible
turbulence, since the crosses in Fig.~\ref{inter} deviate stronger from the K41 prediction than the green line corresponding
to the She--Leveque model. For orders $p=1$ and $2$ of particular interest to this work, corrections $\tau_{p/3}$ appear
to be relatively small.

Our discussion above demonstrates the non-universal nature of scaling relation~(\ref{hb04}), while showing that a 
universal scaling is expected for the fourth-order correlations described by the l.h.s. of equation (\ref{132}) as well as
for a dimensionally equivalent form (\ref{133}). The first- and second-order statistics of $\pmb{v}$ depend on the
relatively small intermittency corrections, $\xi_p=p/3+\tau_{p/3}$ at $p=1,2$. It is hard to tell whether the 
anomalous scaling exponents $\xi_p$ are universal in the limit of infinitely large Reynolds numbers (as $\zeta_p$ 
in the case of incompressible turbulence) or not. Carefully designed large-scale numerical simulations 
will help to improve our understanding of intermittency in supersonic turbulence and address this important question.

\subsection{The turbulent origin of Larson's laws}\label{con}
Is there any direct observational evidence supporting the turbulent origin of the non-thermal velocity fluctuations in MCs?
Do we see any indication that the fundamental relation (\ref{132}) is satisfied? Indeed, the fact that the rate of energy transfer 
per unit volume in the inertial cascade, $\epsilon_{\ell}\equiv\rho_{\ell} u_{\ell}^3/\ell$, is roughly the same in three different components
of the cold neutral medium: in the cold H{\sc i}, in MCs, and in dense molecular cores  \citep{lequeux05}, strongly
suggests that the energy propagates across scales in a universal way consistent with equation (\ref{132}). A more recent compilation
of $^{12}$CO data for the whole population of MCs indicates that $\epsilon_{\ell}$ shows no variation
from structures of $\sim0.01$~pc to GMCs of $\sim100$~pc and has the same value observed in the H{\sc i} gas 
\citep[fig. 6 in][]{hennebelle.12}. Although the dispersion of $\epsilon_{\ell}$ is equally large at all scales, this result 
suggests that MCs traced by $^{12}$CO(1-0) are part of the same turbulent cascade as the atomic ISM.

In the following, we shall exploit the linear scaling of the third-order moment of 
$\delta\bmath v$, which according to equation (\ref{132}) implies a constant kinetic energy density flux within the inertial range.

We shall first derive several secondary 
scaling laws involving the coarse-grained surface density $\Sigma_{\ell}$, assuming that the effective kinetic energy density flux 
is approximately constant within the inertial range (and neglecting gravity). 
Dimensionally, the constant spectral energy flux condition,
\begin{equation}
\rho_{\ell}\left(\delta{}u_{\ell}\right)^3\ell^{-1}\propto\Sigma_{\ell}\left(\delta{}u_{\ell}\right)^3\ell^{-2}\propto\Sigma_{\ell}\ell^{3\zeta_1-2}\approx{}{\rm const.},
\end{equation}
implies
\begin{equation}
\Sigma_{\ell}\propto\ell^{2-3\zeta_1}.
\end{equation}
Substituting $\zeta_1=0.56\pm0.02$, as measured by \citet{heyer.04}, we get $\Sigma_{\ell}\propto\ell^{0.32\pm0.06}$.
We can also rely on the `fractal' properties of the density distribution to evaluate the scaling of $\Sigma_{\ell}$ with $\ell$: 
$\Sigma_{\ell}\propto\rho_{\ell}\ell\propto{}m_{\ell}/\ell^2\propto\ell^{d_{\rm m}-2}$, which in turn implies 
$\Sigma_{\ell}\propto\ell^{0.36\pm0.04}$ for $d_{\rm m}=2.36\pm0.04$ from \citet{romanduval....10}. 
Note that both independent estimates for the scaling of $\Sigma_{\ell}$ 
with $\ell$ agree with each other within one sigma. The observations, thus, indicate that mass surface
density of MCs indeed positively correlates with their size with a scaling exponent $\sim1/3$, which is
consistent with both velocity scaling and self-similar structure of the mass distribution in MCs.

Let us now examine data sets A1 and A2 presented in \citet{heyer...09} for a possible correlation of 
$u_0^{\prime}=\sigma_u{}R^{-1/2}$ with $\Sigma=m/\pi{}R^2$. The A1 data points correspond 
to clouds with their original SRBY87 cloud boundaries, while A2 represents the same set of clouds 
with area within the half-power isophote of the H$_2$ column density. Figure~\ref{sigma} shows the 
data and formal linear least-squares fits with slopes $0.34\pm0.04$ and $0.32\pm0.03$ for A1 (left) 
and A2 (right), respectively. First, note that the Pearson correlation coefficients are relatively small, 
$r=0.524$ for A1 and $r=0.673$ for A2, indicating that the overall quality of the fits is not very high. 
Also note that both correlations are formally not as steep as the virial equilibrium condition~(\ref{vir}) 
would imply. At the same time, when the two data sets are plotted together \citep[as in fig.~7 of][]{heyer...09} 
the apparent shift between A1 and A2 data points caused by the different cloud area definitions in A1 
and A2 creates an {\em impression} of virial equilibrium condition being satisfied. An offset of A1 and A2 
points with respect to the dotted line is interpreted by \citet{heyer...09} as a consequence of LTE-based 
cloud mass underestimating real masses of the sampled clouds. Each of the two data sets, however, 
suggest scaling with a slope around $1/3$ with larger clouds of higher mass surface density being potentially closer 
to virial equilibrium than smaller structures. The same tendency can be traced in the \citet{bolatto....08} 
sample of extragalactic GMCs \citep[][fig.~8]{heyer...09}. A similar trend is recovered by 
\citet{goodman......09} in the L1448 cloud, where a fraction of self-gravitating material obtained from 
dendrogram analysis shows a clear dependence on scale. While most of the emission from the L1448 region
is contained in large-scale `bound' structures, only a low fraction of smaller objects appear self-gravitating.

Let us check a different hypothesis, namely whether the observed scaling $\sigma_u{}R^{-1/2}\propto\Sigma^{1/3}$ is
compatible with the inertial energy cascade phenomenology and with the observed hierarchical structure of MCs. The constant
spectral energy flux condition, $\rho_{\ell}\left(\delta{}u_{\ell}\right)^3\ell^{-1}\approx{}{\rm const.}$, can be recast in terms
of $\Sigma_{\ell}\propto\rho_{\ell}\ell$ assuming $\delta{}u_{\ell}\ell^{-1/2}\propto\Sigma_{\ell}^{\alpha}$ with 
$\alpha\approx1/3$ and $\rho_{\ell}\propto\ell^{d_{\rm m}-3}$,
\begin{equation}
\rho_{\ell}\left(\delta{}u_{\ell}\right)^3\ell^{-1}\propto\rho_{\ell}\Sigma_{\ell}\ell^{3/2}\ell^{-1}\propto\ell^{2(d_{\rm m}-3)+3/2}\approx{}{\rm const.}
\end{equation}
This condition then simply reads as $2(d_{\rm m}-3)+3/2\approx0$ or $d_{\rm m}\approx2.25$, which is
reasonably close to the observed mass dimension.

As we have shown above, the measured correlation of scaling coefficient $u_0^{\prime}$ with the coarse-grained
mass surface density of MCs $\Sigma_{\ell}$ is consistent with a purely turbulent nature of their hierarchical structure. The origin of this
correlation is rooted in highly compressible nature of the turbulence that implies density
dependence of the l.h.s. of equation (\ref{132}). Let us rewrite equation (\ref{133}) for the first-order SF
of the density-weighted velocity: $\langle\left|\delta v_{\ell}\right|\rangle\sim\langle\epsilon_{\ell}^{1/3}\rangle\ell^{1/3}$.
Due to intermittency, the mean cubic root of the dissipation rate is weakly scale-dependent, 
$\langle\epsilon_{\ell}^{1/3}\rangle\propto\ell^{\tau_{1/3}}$, and thus
$\langle\left|\delta v_{\ell}\right|\rangle\propto\ell^{\xi_1}$, where $\xi_1=1/3+\tau_{1/3}$ and $\tau_{1/3}$ 
is the intermittency correction for the dissipation rate. Using dimensional arguments, 
one can write the scaling coefficient in the \citet{heyer...09} relation as
\begin{equation}
\delta{}u_{\ell}\ell^{-1/2}\propto\rho_{\ell}^{-1/3}\ell^{-1/6+\tau_{1/3}}\propto\Sigma_{\ell}^{-1/3}\ell^{1/6+\tau_{1/3}}.
\end{equation}
Since, as we have shown above, $\Sigma_{\ell}\propto\ell^{1/3}$, one gets
\begin{equation}
\delta{}u_{\ell}\ell^{-1/2}\propto\Sigma^{1/6+3\tau_{1/3}}.
\end{equation}
Numerical experiments give $\tau_{1/3}\approx0.055$ for the density-weighted dissipation rate \citep*{pan..09}. 
This value implies $\delta{}u_{\ell}\ell^{-1/2}\propto\Sigma^{0.33}$ consistent with the \citet{heyer...09} 
data.

To summarize, we have shown that none of the three Larson laws is fundamental, but if one of them is known the other two
can be obtained from a single fundamental relation (\ref{132}) that quantifies the energy transfer rate in compressible turbulence.

\section{Gravity and the breakdown of self-similarity}\label{gra}
So far, we have limited the discussion of Larson's relations 
to scales above the sonic scale. Theoretically, the linewidth--size 
scaling index is expected to approach $\zeta_1\approx1/3$ 
at $\ell\lsim\ell_{\rm s}$ in MC substructures not affected by self-gravity \citep[see Section~2 and][]{kritsuk...07a}. 
\citet*{falgarone..09} explored the linewidth--size relation using a large sample of $^{12}$CO 
structures with $\ell\in[10^{-3},10^2]$~pc. These data approximately follow a power law 
$\delta{}u_{\ell}\propto\ell^{1/2}$ for $\ell\gsim1$pc. Although the scatter substantially increases 
below 1~pc, a slope of $1/3$ `is not inconsistent with the data'. $^{12}$CO and $^{13}$CO
observations of translucent clouds indicate that small-scale structures down to a few
hundred astronomical units are possibly intrinsically linked to the formation process of MCs 
\citep*{falgarone..98,heithausen04}.

The observed mass--size scaling index, $d_{\rm m}\approx2.36$, is expected to remain constant 
for non-self-gravitating structures down to $\ell_{\eta}\sim30\eta$, which is about a 
few hundred astronomical units, assuming the Kolmogorov scale $\eta\sim10^{14}$~cm \citep{kritsuk+15.11}. 
This trend is traced down to $\sim0.01$~pc with the recent {\em Herschel} detection of $\sim300$ unbound
starless cores in the Polaris Flare region \citep{andre+10}. For scales below $\ell_{\eta}\sim200$~au, 
in the turbulence dissipation range, numerical experiments predict convergence to $d_m\approx2$, as
the highest density peaks are located in essentially two-dimensional shock-compressed layers, see Fig.~\ref{dm}.

\begin{figure}
\includegraphics[width=84mm]{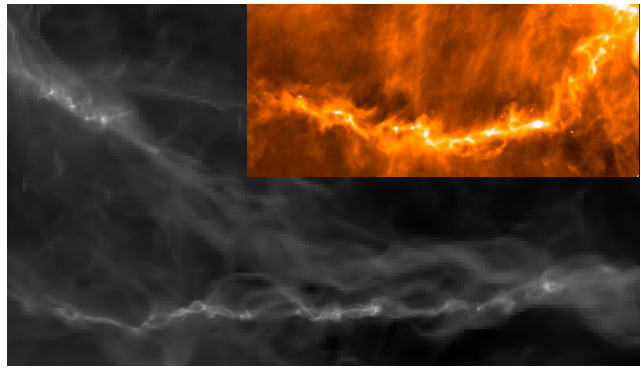}
\caption{The morphology and size scale of self-gravitating filaments that form in a high dynamic range AMR simulation
is similar to that revealed by recent observations. The grey-scale image shows a snapshot of the projected density
distribution in a sub-volume of the (5~pc)$^3$ simulation domain from the self-gravitating model HD3.
The image that shows a region of $\sim3$~pc across was generated using the density field with an effective linear 
resolution of $8192$, while the actual resolution of the simulation was $5\times10^5$.
The colour inset shows {\em Herschel} SPIRE 250~$\mu$m image of the B211/B213 filament in the Taurus MC
\citep{palmeirim+12}, obtained as part of the Herschel Gould Belt survey \citep{andre+10}.}
\label{fil}
\end{figure}

In star-forming clouds, the presence of strongly self-gravitating clumps of high mass surface density
breaks self-similarity imposed by the turbulence. One observational signature of gravity on small scales
is the build-up of a high-end power-law tail in the column density PDF. The tail is observationally associated 
with projected filamentary structures harbouring prestellar cores and young stellar objects \citep{kainulainen...09,andre...11}.

Fig.~\ref{fil} illustrates morphological similarity of the B211/B213 filament in the Taurus MC, as imaged by 
{\em Herschel} SPIRE at 250~$\mu$m, and projected density distribution in a snapshot from the HD3 simulation 
we discussed in Section~\ref{img}. In these high-resolution images, the knotty fragmented structure of dense cores 
sitting in the filament and interconnected by a network of thin bow-like fibres forms a distinct pattern. These
structures could not have formed as a result of monolithic collapse of gravitationally unstable uniform cylinder. 
Supersonic turbulence shapes them by creating non-equilibrium initial conditions for local collapses
\citep{schmidt..12}. Unlike observations, the simulation gives direct access to three-dimensional structure underlying 
what is seen as a filament in projection. Such projected filamentary structures often include chance overlaps of 
physically disjoint segments of an intricate fragmented mass distribution created by the turbulence, see Section~\ref{ms}.

A more quantitative approach to comparing observations and numerical simulations of star-forming clouds deals
with the column density PDFs. Numerical experiments show that the power index $p$ of the high-density tail of the PDF 
[$dS(\Sigma)\propto d(\Sigma^p)$, where $p=-2/(n-1)$] is determined by the density profile [$\rho(r)\propto{}r^{-n}$] 
of a stable attractive self-similar spherical collapse solution appropriate to specific conditions in a turbulent cloud  
\citep{kritsuk..11a}.
For model HD3 that includes self-gravity, we obtained $p\simeq-2.5$ and $n\simeq1.8$ 
in agreement with theoretical predictions. This implies $d_{\rm m}=3-n\simeq1.2$ for the mass--size relation on 
scales $\lsim0.1$~pc. Mapping of the actively star-forming Aquila field with {\em Herschel} gave $p=-2.6\pm0.1$ 
and $d_{\rm m}=1.13\pm0.07$ for a sample of 541 starless cores with size $\ell\in[0.01,0.1]$~pc 
\citep{konyves+10,andre...11,schneider-13}. Using the above formalism, we can predict 
$d_{\rm m}=3-n=2+2/p\simeq1.23$, in reasonable agreement with the direct measurement, given a systematic error
in $p$ due to the variation of opacity with density \citep{schneider-13}. 

Overall, the expected
mass dimension at scales where self-gravity becomes dominant should fall between $d_{\rm m}=1$ 
[Larson-Penston (1969) isothermal collapse solution with $n=2$] and $d_{\rm m}=9/7\simeq1.29$ 
[Penston (1969) pressure-free collapse solution with $n=12/7$; for more detail see \citet{kritsuk..11a}].

The characteristic scale where gravity takes control over from turbulence can be predicted using the 
linewidth--size and mass--size relations discussed in previous sections. Indeed, in a turbulent isothermal gas, 
the coarse-grained Jeans mass is a function of scale $\ell$:
\begin{displaymath}
m^J_{\ell}\propto\sigma_{\ell}^3\rho_{\ell}^{-1/2}\propto\left\{\begin{array}{ll}\!\!\!\rho_{\ell}^{-1/2}\propto\ell^{(3-d_{\rm m})/2}\propto\ell^{0.32},&\!\!\!\!\!\ell\lsim\ell_{\rm s}\\{}\!\!\!\delta v_{\ell}^3\rho_{\ell}^{-3/2}\propto\ell^{1+3(3-d_{\rm m})/2}\propto\ell^{1.96},&\!\!\!\!\!\ell>\ell_{\rm s},\end{array}\right.
\end{displaymath}
where $\sigma_{\ell}^2\equiv\delta{}u_{\ell}^2+c_{\rm s}^2$ \citep{chandrasekhar51} and we assumed that 
$\sigma_{\ell}^2\approx{}c_{\rm s}^2$ at $\ell\lsim\ell_{\rm s}$, while $\sigma_{\ell}^2\approx{}\delta{}u_{\ell}^2$ 
at $\ell\gsim\ell_{\rm s}$. 
The dimensionless stability parameter, $\mu_{\ell}\equiv{}m_{\ell}/m^J_{\ell}$, shows a strong break in slope 
at the sonic scale:
\begin{displaymath}\mu_{\ell}\propto\left\{\begin{array}{ll}\ell^{3(d_{\rm m}-1)/2}\propto\ell^{2.04}&\textrm{if $\ell\lsim\ell_{\rm s}$}\\{}\ell^{(5d_{\rm m}-11)/2}\propto\ell^{0.4}&\textrm{if $\ell>\ell_{\rm s}$,}\end{array}\right.
\end{displaymath}
and a rather mild growth above $\ell_{\rm  s}$. Since both $\mu_{\ell}$ and the free-fall time, 
\begin{equation}
t^{\rm ff}_{\ell}\equiv[3\pi/(32G\rho_{\ell})]^{1/2}\propto\ell^{(3-d_{\rm m})/2}\propto\ell^{0.32},
\end{equation}
correlate positively with $\ell$, a {\em bottom-up} non-linear development of Jeans instability is most likely at $\ell\gsim\ell_{\rm J}$, 
where $\mu_{\ell_{\rm J}}=1$. This assessment is supported by simulations discussed in Section~\ref{img}, which successfully
reproduce observed morphology of filaments (Fig.~\ref{fil}) and density distributions. 
Note that both $\mu_{\ell}$ and
$t^{\rm ff}_{\ell}$ grow approximately linearly (i.e. relatively weakly) with $\Sigma_{\ell}$ at $\ell>\ell_{\rm s}$, while below the sonic
scale $\mu_{\ell}\propto\Sigma_{\ell}^7$. This means that the instability, when present, shuts off very quickly below 
$\ell_{\rm s}$, i.e. $\ell_{\rm J}\sim\ell_{\rm s}$.
The formation of prestellar cores would be possible only in sufficiently overdense regions on scales around 
$\ell_{\rm s}\sim0.1$~pc. The sonic scale, thus, sets the characteristic mass of the core mass function, $m_{\ell_{\rm s}}$, and the
threshold mass surface density for star formation, $\Sigma_{\ell_{\rm s}}$ \citep[cf.][]{krumholz.05,andre+10,padoan.11}.\footnote{%
Our arguments here should be taken with a grain of salt as the Jeans stability criterion we relied on is based on the turbulent 
support concept \citep{chandrasekhar51,bonazzola.....92} originally developed for scales above the turbulent energy 
injection scale, see Section~\ref{intro}. 
In star-forming MCs, the injection scale is believed to be much larger than the sonic length. This illustrates the limits of 
phenomenological approach based on dimensional analysis.}

As supersonic turbulence creates seeds for self-gravitating cores, one can use scaling relations derived above 
to predict the core mass function (CMF). The geometry of turbulence controls the number of overdense 
clumps as a function of their size, $N(\ell)\propto\ell^{-d_{\rm m}}$. The differential size distribution, $dN(\ell)\propto\ell^{-(1+d_{\rm m})}d\ell$, 
together with the marginal stability condition\footnote{$\mu_{\ell}=1$ is a prerequisite for the formation of bound cores from
turbulent clumps based on the \citet{chandrasekhar51} phenomenology.} determine the high-mass end of the CMF. 
Indeed, for $m_{\ell}=m^{\rm J}_{\ell}\propto\ell^{1+3(3-d_{\rm m})/2}$, we obtain a power-law distribution:
\begin{equation}
dN(m)\propto m^{-\alpha}dm,
\end{equation}
where
\begin{equation}
\alpha=(11-d_{\rm m})/(11-3d_{\rm m})\simeq(7+3\zeta_1)/(9\zeta_1-1)
\end{equation}
is reasonably close to Salpeter's index of $2.35$ \citep[cf.][]{padoan.02}.
For instance, if $\zeta_1=\{0.5,0.56\}$, we get $\alpha=\{2.43,2.15\}$, while $d_{\rm m}=\{2.36,2.5\}$ gives 
$\alpha=\{2.20,2.43\}$.

\section{Conclusions and Final Remarks}\label{fin}
We have shown that, with current observational data for large samples of Galactic MCs,
Larson's relations on scales $0.1-50$~pc can be consistently interpreted as an 
empirical signature of supersonic turbulence fed by the large-scale kinetic energy injection. 
Our interpretation is based on a theory of highly compressible turbulence and 
supported by high-resolution numerical simulations. Independently, this picture is corroborated
by often elongated shape of GMCs and their internal filamentary or sheet-like structures \citep[e.g.][]{bally...89,blitz93,molinari+10}.

Our simulations do not rule out the importance of gravitational effects on scales
comparable to the disc scaleheight, where the dynamics become substantially more complex. 
Large-scale gravitational instabilities in galactic discs including both stellar and gaseous 
components \citep[e.g.][]{hoffmann.12} can power a three-dimensional direct energy cascade towards 
smaller scales and a quasi-two-dimensional inverse energy cascade to larger scales 
\citep{elmegreen..01,elmegreen..03,elmegreen...03,bournaud....10,renaud.7.13}. These instabilities can play a
significant role in accumulating the largest and most massive GMCs that appear gravitationally bound 
in observations  \citep[e.g.][]{heyer..01,ballesteros...07,mckee..10,hennebelle.12}.\footnote{The dual cascade
picture, however, narrows the applicability of virial relation (\ref{li}) even further.} Our models implicitly 
include the combined effects of sheared disc instabilities as well as the star formation feedback through 
the external large-scale energy injection that  powers the turbulence. This simplified treatment does not
account for the competition between the large-scale gravity and the feedback, which keeps
the ISM in a self-regulated statistical steady state with GMCs continuously forming, dispersing, and
re-forming \citep{hopkins..12}. The resulting net energy injection rate, however, enters as an input parameter
in our models and can be estimated theoretically \citep{faucher..13} and constrained observationally
(see Section~\ref{mp}).

On small scales, in 
low-density translucent clouds, self-similarity of turbulence can potentially be preserved down to 
$\sim10^{-3}$~pc, where dissipation becomes important. In contrast, in overdense regions, the formation 
of prestellar cores breaks the turbulence-induced scaling and self-gravity assumes control over the slope 
of the mass--size relation. The transition from turbulence- to gravity-dominated regime in this case occurs 
close the sonic scale $\ell_{\rm s}\sim0.1$~pc, where structures turn gravitationally unstable first, leading 
to the formation of prestellar cores.

While some will reason that the virial interpretation 
does not necessarily imply clouds are virialized or even bound and call the acceptance of a turbulent
interpretation of Larson's laws an extreme, others argue that this step solves a number of outstanding problems: it
(i) explains inefficient star formation; (ii) explains the origin and inter-relationship of Larson's laws; (iii)
naturally includes the relationship between GMCs and larger-scale ISM; and finally (iv) it provides a self-consistent
framework for cloud and star formation modelling -- all based on the same set of observational cloud properties.
Moreover, the turbulent interpretation removes other questions previously considered crucial in the isolated cloud 
framework, e.g. whether MC turbulence is driven or decaying, as clouds are born turbulent.

Our approach is based on a theoretical description of turbulent cascade  and the results are valid in a
statistical sense. This means that the scaling relations we discuss hold for sufficiently large ensemble 
averages. Relations obtained for individual MCs and elements of their internal substructure, as well
as relations based on different tracers, can show substantial statistical variations around the mean. 
The scaling exponents we discuss or derive are usually accurate within $\approx(5-10)$ per cent, while
scaling coefficients bear substantial systematic errors. Homogeneous multiscale sampling of a large number 
of MCs and their substructure (including both kinematics and column density mapping) with CCAT, {\em SOFIA}, {\em JWST}
and ALMA will help to detail the emerging picture discussed above.

\section*{Acknowledgements}

We thank Philippe Andr\'e for providing the scaling exponents from {\em Herschel} mapping of 
the Aquila field and for sharing an image of the B211/213 Taurus filament prior to publication. We 
are grateful to Sergey Ustyugov for sharing data from the multiphase simulation we used to 
generate Figs~3 and~4. We appreciate Richard Larson's comments on an early version of the
manuscript, which helped to improve the quality of presentation.
This research was supported in part by NSF grants AST-0808184, AST-0908740, AST-1109570, 
and by XRAC allocation MCA07S014. Simulations and analysis were performed on the DataStar, 
Trestles, and Triton systems at the San Diego Supercomputer Center, UCSD, and on the Kraken 
and Nautilus systems at the National Institute for Computational Science, ORNL.

\bsp

\label{lastpage}

\end{document}